\documentclass[aps,prx,twocolumn,superscriptaddress,amsmath,amssymb,floatfix]{revtex4-2}
\usepackage{graphicx}
\graphicspath{{./figuresSK/}}
\usepackage[usenames,dvipsnames]{color}
\usepackage{siunitx}
\usepackage{upgreek}
\usepackage{float}
\usepackage{tabularx}
\usepackage{bm}
\usepackage[normalem]{ulem}
\usepackage{mathptmx}
\usepackage{ifthen}

\newcommand{\llow}{\ensuremath{l_\text{l}}}
\newcommand{\lup}{\ensuremath{l_\text{u}}}

\newcommand{\Lup}{\ensuremath{L_\text{u}}}
\newcommand{\mlow}{\ensuremath{m_\text{l}}}
\newcommand{\mup}{\ensuremath{m_\text{u}}}
\newcommand{\Omegar}{\ensuremath{\Omega_\text R}}
\newcommand{\omr}{\ensuremath{\Omega_\text{R}}}
\newcommand{\Omegaeff}{\ensuremath{\Omega_\text{eff}}}
\newcommand{\Omegam}{\ensuremath{\Omega}}
\newcommand{\fig}[2]{Fig.\ \ref{#1}\ifthenelse{\equal{#2}{}}{}{(\lowercase{#2})}}
\newcommand{\twofigs}[3]{Figs.\ \ref{#1}(\lowercase{#2}) and \ref{#1}(\lowercase{#3})}
\newcommand{\sect}[1]{Sec.\ \ref{#1}}
\newcommand{\eq}[1]{Eq.\ \eqref{#1}}

\renewcommand{\paragraph}[1]{{\ignorespaces\bigskip\noindent\textbf{\textsf{#1}}\newline}}
\renewcommand{\subparagraph}[1]{{\medskip\noindent\textbf{#1.}\ }}

\newcommand{\affiliationPDI}{\affiliation{Paul-Drude-Institut f\"ur Festk\"orperelektronik (PDI), Leibniz-Institut im Forschungsverbund Berlin e.V.,
Hausvogteiplatz 5-7, 10117 Berlin, Germany}}
\newcommand{\affiliationLMU}{\affiliation{Center for NanoScience (CeNS) \& Fakult\"at f\"ur Physik, Ludwig-Maximilians-Universit\"at (LMU), 80539 M\"unchen, Germany}}
\newcommand{\affiliationCSIC}{\affiliation{Instituto de Ciencia de Materiales de Madrid, CSIC, 28049 Madrid, Spain}}
\newcommand{\affiliationICTP}{\affiliation{The Abdus Salam International Centre for Theoretical Physics (ICTP), Strada Costiera 11, 34151 Trieste, Italy}}
\newcommand{\affiliationIBS}{\affiliation{Center for Theoretical Physics of Complex Systems (PCS), Institute for Basic Science (IBS), Expo-ro 55, Yuseong-gu, Daejeon 34126, Korea}}

\begin{document}

\title{Visualized Wave Mechanics by Coupled Macroscopic Pendula: Classical Analogue to Driven Quantum Bits} 

\author{Heribert Lorenz}\affiliationLMU

\author{Sigmund Kohler}\affiliationCSIC

\author{Anton Parafilo}\affiliationIBS

\author{Mikhail Kiselev}\affiliationICTP

\author{Stefan Ludwig}\affiliationPDI

\begin{abstract}
Quantum mechanics increasingly penetrates modern technologies but, due to its non-deterministic nature seemingly contradicting our classical everyday world, our comprehension often stays elusive. Arguing along the correspondence principle, classical mechanics is often seen as a theory for large systems where quantum coherence is completely averaged out.  Surprisingly, it is still possible to reconstruct the coherent dynamics of a quantum bit (qubit) by using a classical model system. This classical-to-quantum analogue is based on wave mechanics, which applies to both, the classical and the quantum world. In this spirit we investigate the dynamics of macroscopic physical pendula with a modulated coupling. As a proof of principle, we demonstrate full control of our one-to-one analogue to a qubit by realizing Rabi oscillations, Landau-Zener (LZ) transitions and Landau-Zener-St\"uckelberg-Majorana (LZSM) interferometry. Our classical qubit demonstrator can help comprehending and developing useful quantum technologies.
\end{abstract}

\date{\today}

\maketitle


%
Quantum technology already has a drastic impact on society. This development presently accelerates with our growing ability to harvest coherent quantum dynamics for engineering game changing devices such as quantum computers or a quantum internet. At the same time, while the mathematical framework of quantum mechanics can be considered complete, fundamental aspects of the underlying physics, even on the level of only few qubits are outside our empirical world. In this situation, classical model systems capable of enlightening the often elusive coherent dynamics of quantum systems may prove very useful \cite{ShoreAJP09}. This approach might be fundamentally questioned due to a central paradigm of quantum dynamics, which is its probabilistic nature in contrast to the deterministic classical equation of motion (EOM). Nevertheless, besides non-determinism and non-locality, wave properties and the superposition principle being central elements of quantum mechanics appear also in classical physics. For example, the quantum mechanical double split experiment may be visualized with classical water waves. Here we visualize one of the most basic quantum systems, a qubit, by physical macroscopic pendula.

Classical dynamics can generally be formulated in terms of second-order, non-linear and inhomogeneous differential equations, while non relativistic quantum mechanics is based on the first order, homogeneous and linear Schr\"odinger equation. Hence, most classical systems are improper for simulating qubit dynamics. In this article, we derive the conditions under which classical pendula with modulated coupling nevertheless can be described by a Schr\"odinger-like equation. We demonstrate this classical-to-quantum analogue by exploring three realizations of qubit control, namely Rabi oscillations \cite{RabiPR37}, LZ transitions \cite{LandauPZS32a,ZenerPRSLA32} and, finally, LZSM interferometry \cite{StueckelbergHPA32,MajoranaNC32}. 

Recent developments in quantum technology have motivated theoretical \cite{ShoreAJP09,GronbechJensenPRL05,NovotnyAJP10,HeinrichPRA10,FrimmerAJP14,IvakhnenkoSR18,ParafiloPRB18} and experimental \cite{SusstrunkS15,NashPNAS15,FaustNP13,SeitnerPRB16} projects exploring analogues between classical coupled oscillators and its quantum version. The most interesting dynamics happens at avoided crossings of the eigenmodes of coupled oscillators near resonance. Previous theoretical considerations \cite{NovotnyAJP10,FrimmerAJP14,IvakhnenkoSR18} and experiments \cite{FaustNP13,SeitnerPRB16} with nanomechanical oscillators used a time-dependent frequency difference corresponding to the detuning usually modulated in case of qubits \cite{MullenPRL89,VitanovPRA99,WubsNJP05,SillanpaaPRL05,BernsNL08,StehlikPRB12,ForsterPRL14,ForsterPRB15b,HeinrichNN21}. To experimentally establish a classical-to-quantum analogue based on macroscopic pendula we instead modulate the coupling, which for this system is more practical than driving the detuning. This gimmick, for the first time allows us to continuously monitor the coherent dynamics of a driven two-level system at ambient conditions and to observe it with bare eyes. As we establish a one-to-one correspondence, our coupled pendula directly visualize the coherent dynamics of a driven qubit.

\paragraph{Setup and model}
\begin{figure}[tb]
\centering
\includegraphics{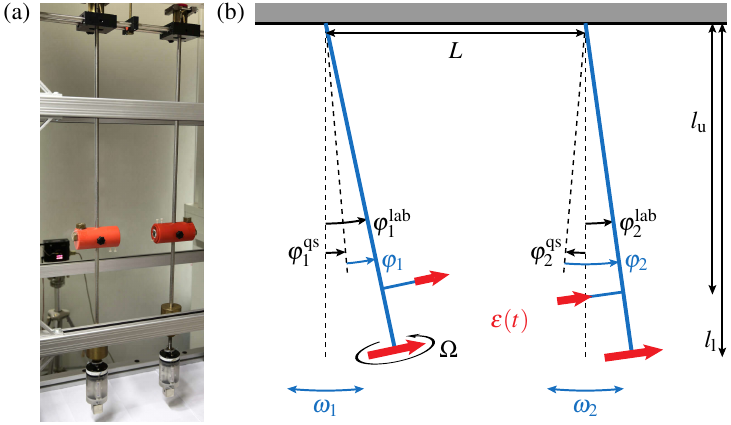}
\caption{Photograph (a) and sketch (b) of the pendula coupled via cubic neodymium magnets (red arrows in sketch indicate magnetic moments), a lower pair with moments $\mlow=25.37\,\text{Am}^2$ and an optional upper pair with $\mup=6.54\,\text{Am}^2$. The lower magnets are attached at the end of the pendula while the upper ones sit inside the red cylinders. Their respective distances from the pivots are $\llow=1.148\,$m versus $\lup=0.635\,$m. One of the lower magnets is slowly rotated at angular frequency $\Omega$ around the pendulum rod by a battery driven motor (inside the transparent plastic cases). The distances $L$ between the pivots and $\Lup$ between the upper magnets at deflections $\varphi^\text{lab}_1=\varphi^\text{lab}_2=0$ as well as $\Omega$ are variable. Each pendulum weighs 4.242\,kg, where a 2.1\,kg brass weight (visible in the photo) can be moved along a threaded section of each rod to vary $\omega_1$ and $\omega_2$.
}
\label{fig:model}
\end{figure}%
In \fig{fig:model}{} we display a photograph and a simplified sketch of the setup. It consists of two pendula, each being described by its deflection angle $\varphi^\text{lab}_k$ and its angular frequency $\omega_k$ with $k=1,2$. The two pendula are coupled via permanent magnets and detuned by the frequency difference $\Delta=\omega_1-\omega_2$. To probe the dynamics of qubits, usually the energy detuning between the diabatic states is modulated. However, modulating the coupling is mathematically equivalent after applying the appropriate basis transformation. For our system it is more practical to modulate the coupling. For this purpose we employ a battery driven linear motor, which rotates one of the magnets around the axis defined by the pendulum rod. As a result, the coupling and, at the same time, the equilibrium deflections of the pendula, $\varphi_k^\text{qs}$, are periodically modulated in time. The latter correspond to the quasistatic solution of the driven system, describing the (momentary) adiabatic equilibrium position.

We consider the deviation from the adiabatic equilibrium, $\varphi_k = \varphi^\text{lab}_k - \varphi_k^\text{qs}$. Aiming at a description in the form of a Schr\"odinger equation, which is linear and of first order, experiments and theory have to facilitate a linearization of the non-linear Newton EOM. This requires small deflection angles, small frequency differences and similar moments of inertia of the uncoupled pendula. The linearized version of the EOM reads
\begin{equation}
\begin{split}
\ddot\varphi_1 + \omega_1^2\varphi_1
={} \omega_0\varepsilon(t) (\varphi_1-\varphi_2) ,
\\
\ddot\varphi_2 + \omega_2^2\varphi_2
={} \omega_0\varepsilon(t) (\varphi_2-\varphi_1) ,
\end{split}
\label{newton}
\end{equation}
where $\omega_0=\frac{1}{2}(\omega_1+\omega_2)$ is the average pendulum (angular) frequency and $\varepsilon(t)$ is the coupling in units of frequency \footnote{In the interaction term, we have neglected the small difference of the moments of interia.  Moreover, the sign of $\varepsilon(t)$ is chosen such that it matches the usual definition in the quantum mechanical two-level problem. It is positive for attractive interaction.}. The symmetry of the interaction terms on the right-hand side of \eq{newton} is essential for resembling the Schr\"odinger equation and requires neglecting the difference between the moments of inertia.  Our modulated coupling, $\varepsilon(t)$, corresponds to the time-dependent level detuning commonly used to drive qubits, for instance in the context of the quantum mechanical LZSM problem \cite{ShevchenkoPR10, IvakhnenkoPR23}. To simplify a comparison with typical qubit experiments we aim at a coupling of the common form $\varepsilon(t) = \varepsilon_0 + A\cos(\Omega t)$, which renders \eq{newton} a Mathieu equation \cite{MathieuJMPA68}. This modulation requires an experimental setup allowing for the far-field approximation of the dipole-dipole interaction. 

For simulating qubit experiments we would like to independently modify the mean coupling $\varepsilon_0$ and the modulation amplitude $A$. To achieve this, we use two sets of magnet pairs, see \fig{fig:model}{}. The lower magnets are attached at the distance \llow\ from the pivots and the upper magnets at \lup, where $(\llow-\lup)/L$ is sufficiently large to allow us to neglect quadrupole components of the coupling. The coupling is then composed of the sum of the contributions of the upper versus lower magnets, $\varepsilon=\varepsilon_\text{u}+\varepsilon_\text{l}$, where we slowly modulate $\varepsilon_\text{l}$ by rotating one of the lower magnets. The time-dependence of the reference point of the linearization, $\varphi_k^\text{qs}(t)$, leads to harmonic mixing such that $\varepsilon_0$ aquires a contribution from the rotating magnets and, vice versa, $A$ is also affected by the static magnets.

The linearized EOM \eq{newton}, which is still of second order, describes the free oscillations of two pendula with modulated coupling. In comparison, the Schr\"odinger equation of a qubit describes probability functions. These correspond to the slowly varying occupation amplitudes of the two pendula, given by the envelope functions, say $\Psi_k$, of the individual rapid oscillations $\varphi_k(t)$. To separate the time scales, we therefore employ the ansatz
\begin{equation}
\varphi_{k} = e^{-i\omega_0t} \Psi_{k} + \text{c.c.}
\label{ansatz}
\end{equation}
with a rapidly oscillating prefactor and slowly varying complex envelopes $\Psi_{k}$. Inserting the ansatz into Eq.~\eqref{newton}, while neglecting second order derivatives of $\Psi_k$, we find
\begin{equation}\label{Schroedinger_Rabi}
i\frac{d}{dt}\begin{pmatrix} \Psi_1 \\ \Psi_2 \end{pmatrix}
= \frac{1}{2}\begin{pmatrix}
    \Delta -\varepsilon(t) & \varepsilon(t) \\
    \varepsilon(t) & -\Delta -\varepsilon(t) \end{pmatrix}
   \begin{pmatrix} \Psi_1 \\ \Psi_2 \end{pmatrix},
\end{equation}
which for $\hbar=1$ possesses the form of a Schr\"odinger equation of the driven two-level system in the representation frequently found in textbooks for the Rabi problem (in a gauge without $\varepsilon$ in the diagonal).

For describing LZ transitions of a qubit with time-dependent detuning one usually uses the diabatic basis, in which the constant tunnel coupling appears in the off-diagonal matrix elements of the Hamiltonian. As we modulate the coupling between our pendula, it is convenient to transform into the according diabatic basis of the in-phase and out-of-phase modes, $\varphi_\pm = (\varphi_1\pm\varphi_2)/2$, in which the constant frequency difference $\Delta$ appears in the off-diagonal elements of the Hamiltonian.  With $\Psi_\pm$ defined in accordance with Eq.~\eqref{ansatz} the presentation of the LZ problem for our pendula then reads
\begin{equation}\label{Schroedinger_LZ}
i\frac{d}{dt}\begin{pmatrix} \Psi_+ \\ \Psi_- \end{pmatrix}
=
\frac{1}{2}
\begin{pmatrix}
   0 & \Delta \\
   \Delta & -2\varepsilon(t)
\end{pmatrix} \begin{pmatrix} \Psi_+ \\ \Psi_- \end{pmatrix}\,.
\end{equation}

Equations \eqref{Schroedinger_Rabi} and \eqref{Schroedinger_LZ} provide the foundation for comparing the dynamics of classical pendula with that of a qubit. They describe the occupation amplitudes of our coupled pendula in the form of a Schr\"odinger equation in the two alternative bases $\{\varphi_1,\varphi_2\}$ versus $\{\varphi_+,\varphi_-\}$. In Appendices A and B we offer an elegant alternative derivation based on the Lagrange formulation of classical mechanics and starting from the non-linear Newton equation. We also demonstrate there, how the time-dependent quasi static equilibrium, $\varphi^\text{qs}_k(t)$, contributes to the static $\varepsilon_0$ and discuss limitations of our approximations.

In a nutshell, \eq{Schroedinger_Rabi} describes ---for the case of a time dependent coupling between the pendula--- the dynamics of the occupation amplitudes of the individual pendula. In this way, the eigenmodes of the uncoupled pendula directly correspond to the wave functions of the localized sates of a qubit. Equation (\ref{Schroedinger_Rabi}) is the natural choice for predicting Rabi oscillations occurring between the occupation amplitudes of the individual pendula for weak coupling. A basis transformation yields \eq{Schroedinger_LZ}, which describes the dynamics between the occupation amplitudes of the in-phase and out-of-phase superposition modes of the individual pendula. Without driving, they resemble the eigenmodes of two strongly coupled pendula and correspond to the eigenfunctions of a qubit at zero detuning. Consequently, \eq{Schroedinger_LZ} is the natural choice for predicting the dynamics of the occupation amplitudes in the regime of LZ transitions, where the maximum coupling exceeds the detuning. See Appendix \ref{sec:duality} and Table \ref{table:params} therein for a one-to-one comparison between the parameters of a qubit and the pendula.

In our experiments we control $A$ and $\varepsilon_0$ by varying the distance $L$ between the lower magnets, corresponding to the distance between the pivots, and optionally the distance \Lup\ between the upper magnets [positioned inside the red horizontal cylindric housings seen in \fig{fig:model}a], both defined for $\varphi_1=\varphi_2=0$. In addition, we adjust the frequency difference, $\Delta$, by moving a heavy weight [vertical brass cylinders partly visible in \fig{fig:model}a] using a standard thread along one of the pendulum rods. We employ a line scan camera to simultaneously image at a rate of 10\,Hz the lateral positions of both pendula within a linear pixel array. Applying numerical filtering we then obtain the displacement angles $\varphi_{k}(t)$ as a function of time, see Appendix \ref{app:experiment}. The mean frequency of our pendula is close to $\omega_0/2\pi\simeq 0.5$\,Hz. To ensure the validity of the linearized \eq{newton}, we work with small deflections $|\varphi_k|<1^\circ$, modulation frequencies $\Omegam<10^{-2}\omega_0$ and frequency differences $|\Delta|<0.1 \omega_0$. Thereby $\omega_0$ is always the largest frequency by far supporting the separation of time scales in \eq{ansatz}. The quality factor of $Q>2500$ of the coupled pendula is high enough to allow us ignoring dissipation as we did in the model above. In order to achieve high and stable quality factors, we employ professional pendulum clock pivots based on leaf springs provided by the company Erwin Sattler GmbH \& Co. KG. In detail our experimental results indicate, that the damping of the coupled pendula motion is dominated by magnetic induction, i.e., eddy currents induced inside the conducting magnets due to their relative motion. The friction of air and the deforming pivot springs dominate the damping of the uncoupled pendula with stable quality factors of $Q>10000$. Note, that we can easily achieve the strong coupling regime as our maximal coupling of $|\varepsilon|_\text{max}\simeq0.2$\,Hz exceeds the resonance line width of $\omega_0/2\pi Q\sim10^{-4}\,$Hz by more than three orders of magnitude.

\paragraph{Analysis and Discussion}
To explore the analogy between our coupled pendula and a qubit we first perform Rabi oscillations between the individual pendula in the limit of fast driving  with $\Omegam\simeq|\Delta|\gg A$, where the driving amplitude becomes the Rabi frequency, $\Omegar\simeq A$. After that, we turn to ``qubit manipulation'' using LZ transitions between the superposition modes, $\varphi_\pm$, in the limit of slow driving, $\Omegam\ll|\Delta|<A$. 

\begin{figure}[t]
\centering
\includegraphics{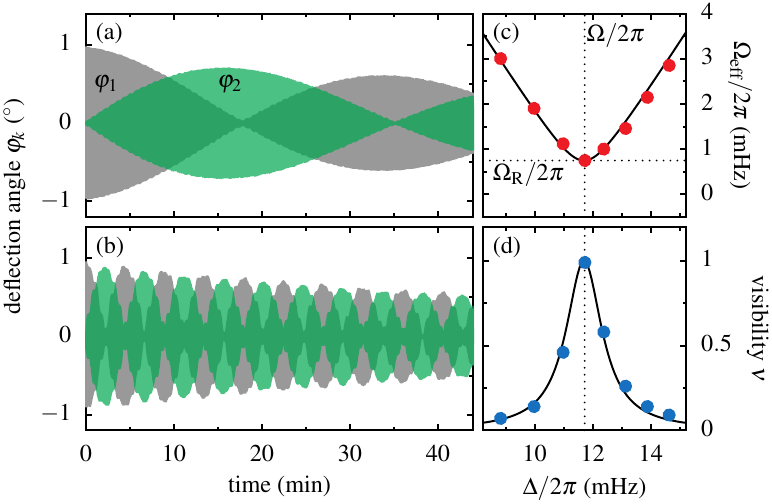}
\caption{
Near resonance Rabi oscillations between the two pendula with mean frequency $\omega_0/2\pi\simeq528$\,mHz, frequency difference $\Delta/2\pi=11.7\,$mHz and modulation frequency $\Omegam/2\pi=11.8\,$mHz. At $t=0$ pendulum 1 was deflected at maximally attracting lower and no upper magnets. Individual oscillations are not visible owing to the time axis covering 45 minutes. (a,b) Deflections $\varphi_1(t)$ and $\varphi_2(t)$ of the two pendula for the pivot distances $L=496.5\,$mm and $L=330.0\,$mm resulting in Rabi frequencies of $\Omegar/2\pi=0.47\,$mHz versus $\Omegar/2\pi=3.69\,$mHz. (c,d) Effective frequency $\Omegaeff(\Delta)$ and visibility $\nu(\Delta)$ of the Rabi oscillations for $L=454.0\,$mm. The solid lines represent model predictions.
}
\label{fig:rabi}
\end{figure}

In Figs.\ \ref{fig:rabi}{a} and \ref{fig:rabi}{b}
we present Rabi oscillations for two different pivot distances $L$ but otherwise identical conditions. We use no upper magnets and weak couplings (large $L$), such that $\varepsilon_0\ll A$ and $\varepsilon(t)\simeq A\cos(\Omega t)$. Shown are the deflections $\varphi_k(t)$ in respect to the equilibrium $\varphi_k^\text{qs}$. The observed beatings between the pendula are Rabi oscillations, where the variation between Rabi frequencies in \fig{fig:rabi}{a} versus \fig{fig:rabi}{b} reflect the differences in $L$. In both cases the energy transfer between the pendula is almost complete, as we chose a near resonance condition $\Omegam\simeq|\Delta|$.  Small steps, which occur at the repetition rate $2\Omega$ [zoom into  \twofigs{fig:rabi}{a}{b} to clearly see them], indicate side band transitions beyond the rotating wave approximation. In \fig{fig:rabi}{b} these steps are bigger due to a larger modulation amplitude compared to \fig{fig:rabi}{a}.

By varying $\Delta$ we next explore the Rabi dynamics near resonance. In \fig{fig:rabi}{c} we present the effective Rabi frequency $\Omegaeff(\Delta)$ corresponding to the actual beating frequency. Likewise, in \fig{fig:rabi}{d} we show the visibility $\nu(\Delta)$ defined as the fraction of energy exchanged between the pendula. Symbols are measured data while the lines visualize the theory predictions, $\Omegaeff=[\Omegar^2+(\Omegam-|\Delta|)^2]^{1/2}$ and $\nu=\Omegar^2/\Omegaeff^2$. The only fit parameter is the Rabi frequency $\Omegar/2\pi=0.71\,$mHz, which defines the minimum of $\Omegaeff$ at resonance $\Omegam=\Delta$ and which can be used to accurately determine the magnetic moment \mlow, see Appendix \ref{app:drivenqubit}. The excellent agreement between theory and experiment underlines the high quality of our classical mechanics experiment. Since the model curves can be derived from the Schr\"odinger equation \eqref{Schroedinger_Rabi}, the result establishes a first analogue between classical pendula and a qubit.      

\begin{figure}[tb]
\centering
\includegraphics{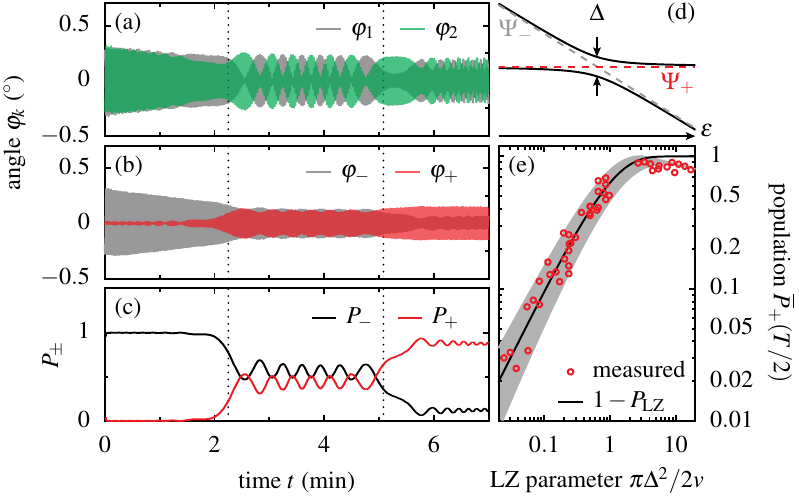}
\caption{LZ transitions:
At $t=0$ both pendula where deflected to excite the out-of-phase mode $\varphi_-$ at maximally attracting lower and no upper magnets.
(a--c) Measured deflection angles $\varphi_{1,2}(t)$ (adiabatic modes) in a, diabatic modes $\varphi_\pm(t)$ in b and their occupations $P_\pm(t)$ in c during the first period $T=2\pi/\Omega$ of modulation for $\Omegam/2\pi=2.27\,$mHz, $\Delta/2\pi=6.7\,$mHz, $\omega_0/2\pi=0.53$\,Hz, and $P_\text{LZ}\simeq0.4$. Two LZ transitions occur as the pendula pass through avoided crossings at $\varepsilon=0$ indicated by vertical dotted lines. The occupations $P_\pm(t)$ as well as the beating dynamics clearly change at each avoided crossing.
(d) Sketch of the avoided crossing. Solid lines are the eigenfrequencies $(\mp\sqrt{\Delta^2+\varepsilon^2}-\varepsilon)/2$ and dashed lines indicate the frequencies of the envelope functions $\Psi_\pm(\varepsilon)$.
(e) Mean population $\overline{P}_+(T/2)$ averaged over a proper time window around $t=T/2$ in between the first two crossings. Both, $\Delta$ and the LZ speed $v$ are varied between individual measurements. The black line follows $1-P_\text{LZ}$. Our initialization at finite $\varepsilon(t=0)$ causes a small initial population of the upper mode $\varphi_+$, which varies from measurement to measurement in amplitude and phase. The gray region indicates the corrected range of prediction accounting for the range of experimental parameters by assuming arbitrary initial phases.
}
\label{fig:LZ}
\end{figure}

An elegant method to manipulate qubits in the limit of slow modulation, $|\Delta|\gg\Omegam$, are LZ transitions \cite{SaitoEL06, RibeiroPRL09}.  In \fig{fig:LZ}{a}
we present the deflection angles $\varphi_{k}$, while one of the magnets completes one full rotation. Within this driving period, the pendula pass twice through the avoided crossing at zero coupling, sketched in \fig{fig:LZ}{d}, namely from positive to negative $\varepsilon$ and back.  At $t=0$, we initialized $\varphi_{1}=-\varphi_{2}$, such that $\varphi_+=0$. This is evident in \fig{fig:LZ}{b} plotting the in-phase and out-of-phase combinations $\varphi_\pm=(\varphi_1\pm\varphi_2)/2$. In \fig{fig:LZ}{c} we present the according populations $P_\pm$, given by the square modulus of the slowly varying amplitude, $P_\pm\propto|\Psi_\pm|^2$, normalized such that $P_++P_-=1$. Around the two avoided crossings (indicated by vertical dotted lines) we observe LZ transitions. The first one, which mixes $\varphi_+$ and $\varphi_-$, is followed by beats with the time dependent frequency $\sqrt{\Delta^2+\varepsilon^2(t)}$ clearly visible between $\varphi_{1}$ and $\varphi_{2}$ as well as between $P_-$ and $P_+$. The latter beats confirm theoretical predictions \cite{MullenPRL89,VitanovPRA99,WubsNJP05}, namely chirped oscillations centered around the LZ probability $P_\text{LZ} = \exp(-\pi\Delta^2/2v)$ and $1-P_\text{LZ}$, respectively, right after passing through the avoided crossing. Here, $v = \Omega(A^2-\varepsilon_0^2)^{1/2}$ is the sweep velocity at $\varepsilon=0$, which depends on  $L$, \Lup\ and $\Omegam$. This observation demonstrates an advantage of our classical two-level system, which---in contrast to a qubit---allows us to trace the time evolution of population probabilities in real time in a single shot measurement. Based on the comparison with theory, we identify the long-time transition probability of a single LZ transition by averaging out the beats of the measured occupation $P_+(t)$ within an appropriate time window after half of the modulation period $T=2\pi/\Omega$, i.e., centered between the two LZ transitions. In \fig{fig:LZ}{e} we compare the resulting $\overline P_+(T/2)$ for a wide range of the parameters $\Delta$, $L$, \Lup\ and $\Omegam$ with the classic result $1-P_\text{LZ}$ for the limit $t\to\infty$ \cite{LandauPZS32a, ZenerPRSLA32, StueckelbergHPA32, MajoranaNC32}.

The agreement between model and measurements is good albeit compared to the Rabi experiment above the data scatter considerably around the model line. These deviations can be explained with the initialization into $\varphi_-$ at finite $\varepsilon$, which for $\Delta\ne 0$ is not an eigenmode, as it would be for an initialization at $\varepsilon\to-\infty$ \cite{LandauPZS32a, ZenerPRSLA32}. The weak admixture of the second eigenmode gives rise to a weak beating between $\varphi_\pm(t)$ right after initialization as visible in \fig{fig:LZ}b for $t\lesssim1\,$min. Treating the relative phase between the modes $\varphi_\pm(t)$ (which could be predetermined at the cost of additional experimental effort) as an unknown, we predict the range of possible values of $\overline P_+(\pi/\Omega)$ for arbitrary phases [gray area in \fig{fig:LZ}{e}]. In the adiabatic limit, $\Delta^2/v\gg1$, independent of the relative phase the finite occupation of the upper eigenmode, $P_+(t=0)>0$, results in $\overline P_+(\pi/\Omega)<1$ while $P_\text{LZ}=1$.

Note, that in a corresponding experiment with an actual qubit the initialization procedure would be similar, such that the phase problem described above occurs as well. However, only the classical qubit analog allowed us to perform continuous measurements as those in \fig{fig:LZ}{c}, which helped us to fully determine the influence of the non-zero phase at initialization.
This result is an example of the usefulness of our classical approach for deciphering the sometimes complex dynamics of qubits.

\begin{figure}[tb]
\centering
\includegraphics{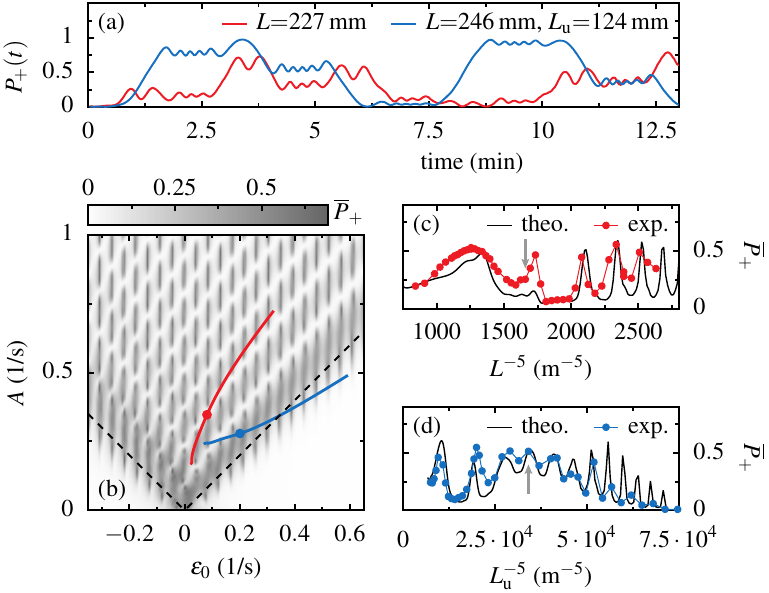}
\caption{LZSM interference:
(a) Measured occupation $P_+(t)$ of the in-phase mode $\varphi_+$, while one magnet was rotated at $\Omega=\SI{7.1}{mHz}$ with (blue) and without (red) static magnets. At $t=0$, $\varphi_-$ was excited as in Fig.\ \ref {fig:LZ}.
(b) Occupation $P_+$ averaged over the first 5
driving periods as a function of the average detunging $\epsilon_0$ and the
effective driving amplitude $A$. The data are computed using \eq{Schroedinger_LZ}.
(c,d) Slices through the LZSM fan along the lines marked in panel b.
Dots correspond to experimental results, while solid black lines show
theory data from panel b. The colored dots in panel b and the arrows in panels c and d indicate the values used in panel a.
}
\label{fig:LZSM}
\end{figure}

Each LZ transition mixes the eigenmodes as demonstrated in \fig{fig:LZ}{e}. The resulting superposition state then accumulates the adiabatic phase $\int \text{d}t (\Delta^2+\varepsilon(t)^2)^{1/2}$ integrating the difference between the two eigenfrequencies with the Stokes phase added \cite{ShevchenkoPR10}. The second LZ transition is then heavily influenced by these phases \cite{StueckelbergHPA32, MajoranaNC32}. Indeed, multiple LZ transitions result in a complicated time evolution of $P_+(t)$ as can be seen in \fig{fig:LZSM}a,
which presents two example time traces of $P_+(t)$. While the modulation frequency is identical for both measurements,  $\Omega/2\pi=7.1\,$mHz, we varied $A$ and $\varepsilon_0$. Each trace covers five modulation periods corresponding to ten LZ transitions. Most of them are clearly visible as more or less pronounced steps while for some transitions $P_+(t)$ stays almost unchanged beyond the perpetual beating. The steady state solution for continuous driving averaging $P_+(t)$ over many periods gives rise to LZSM interference patterns $\overline P_+(A,\varepsilon_0)$, which can be used for exploring qubit dynamics, decoherence or multi-color driving \cite{ForsterPRL14,ForsterPRB15b}. For a practical comparison, allowing for small deviations, we average $P_+(t)$ over the initial five modulation periods. In \fig{fig:LZSM}{b} we present the LZSM interference pattern $\overline P_+(A,\varepsilon_0)$ as a gray scale, which we computed using the Schr\"odinger equation \eqref{Schroedinger_LZ}. 
In \twofigs{fig:LZSM}{c}{d} we finally present interference traces along the two solid lines in \fig{fig:LZSM}b, mind the color code. In \fig{fig:LZSM}{c} we plot $P_+(L^5)$ for the case without upper magnets, while for \fig{fig:LZSM}{d} we added the upper magnets and show $P_+(L_\text{u}^5)$ for a constant pivot distance, $L=246$\,mm. Solid lines are model curves calculated with \eq{Schroedinger_LZ} [contained in the gray scale plot in \fig{fig:LZSM}b], while the dots present measured interference patterns. Hereby, each point corresponds to the average of a $P_+(t)$ trace as those shown in \fig{fig:LZSM}{a}. Our measurements qualitatively reproduce the calculated interference fringes. Quantitative deviations indicate the limitations of the mapping of the Newton equation onto the Schr\"odinger equation, in particular for large $A$ and $\epsilon_0$. In Appendix \ref{app:dataanalysis} we provide related background information.

\paragraph{Summary}
Wave mechanics introduced by Erwin Schr\"odinger in 1926 \cite{SchrodingerPR26} provides a mathematical description for the coherent dynamics of a qubit. However, a continuous experimental visualization of its time evolution is hindered by the principle of projection measurement; each measurement would destroy the quantum coherence. In comparison, our classical qubit analog surely allows to trace the complete time evolution in a single measurement. In order to actually visualize Schr\"odingers wave mechanics using physical pendula with modulated coupling one has to map the non-linear, second order and inhomogeneous classical EOM to the linear, first order and homogeneous Schr\"odinger equation. This mapping, which includes a linearization, a rotating wave approximation and a time-dependent shift of the reference point, also clarifies the experimental conditions necessary for classical qubit simulator experiments and their physical interpretations. In this spirit, we presented three key qubit experiments with coupled pendula, namely Rabi oscillations, LZ transitions and LZSM interferometry. Comparing measurements with the prediction of the Schr\"odinger equation we demonstrated that our classical experiments directly visualize Schr\"odinger's wave mechanics. Our classical qubit simulator bridges the gap between the elusiveness of quantum mechanics and the common imagination pre-shaped by classical experiences. The experimental setup is highly versatile and might be used for exploring a variety of phenomena beyond simulating a qubit, such as geometrical phases \cite{BerryPRSA84}, after adding more pendula, multi-level systems \cite{MiPRB18,ShevchenkoPRB18} or coherent transfer by adiabatic passage \cite{MenchonEnrichRPP16} or simulating the non-linear Schr\"odinger equation \cite{LeggettRMP01}. Moreover, driven systems of coupled pendula may serve as visualizer of a large variety of coupled systems in nature or even economical, social or financial systems. However, a limit to any classical model system is imposed by entanglement of multiple qubits, which has no classical analog.

\paragraph{Acknowledgements}
The authors thank W.\ Kurpas and S.\ Manus for technical support. H.L.\ thanks K.\ Pfeffer and R.\ Buchholz for frequent discussions of the ongoing experiments. This work was financially supported by the Center for NanoScience (CeNS) at LMU Munich, by the Spanish Ministry of Science and Innovation through Grant.\ No.\ PID2020-117787GB-I00, the CSIC Research Platform on Quantum Technologies PTI-001 and the Institute for Basic Science in Korea (IBS-R024-D1). The work of M.K.\ is conducted within the framework of the Trieste Institute for Theoretical Quantum Technologies (TQT).

\paragraph{Author's contributions}
S.L. and M.K. initiated the project and planned it together with H.L. The experimental setup of the coupled pendula was designed by H.L. and S.L\@. H.L. performed the experiments. The data were analyzed by all authors. M.K. and S.K. formulated the theoretical model. S.K., M.K., and A.P. computed the theoretical data. S.K., M.K. and S.L. wrote the article including the appendices.


\clearpage

\appendix
\section{Newton equation}
\label{app:newton}

\begin{figure}[b]
\begin{center}
\includegraphics{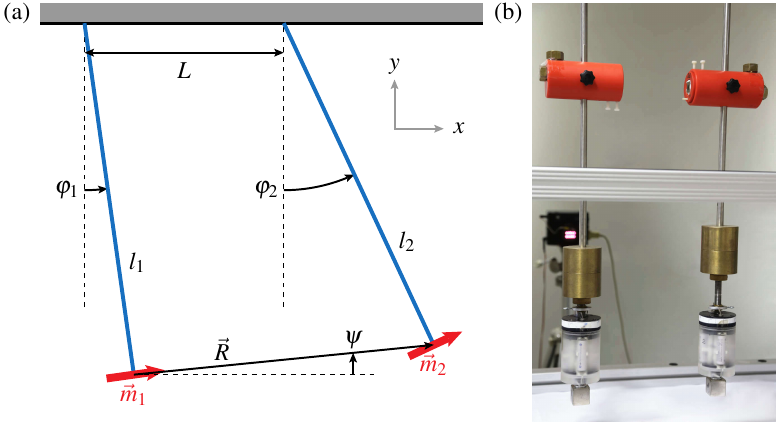}
\caption{Sketch (a) and photograph (b) of the pendula, which are coupled by cubic neodymium magnets. The sketch visualizes the geometry of the pendula, the orientations of the magnets as well as the angles and lengths used in the derivation of the interaction potential. The two lower magnets with magnetizations $\vec{m}_1$ and $\vec{m}_2$ (with $|\vec{m}_1|=|\vec{m}_2|$) are attached at the end of the pendula to motors, which are encapsulated together with batteries and a circuit board by half transparent polethylene cylinders.The upper magnets, which have an approximately four times smaller magnetization, can be horizontally moved within the red ceramic cylinders in order to adjust their distance independently from the distance $L$ between the pivots. With their magnetizations put parallel and in line they attract each other. Brass nuts attached to the cylinders have the purpose of balancing the weights of the magnets such that the center of masses stay within the pendulum rods. The 2.1\,kg heavy brass weights above the motors can be vertically moved to tune the pendula's eigenfrequencies. The line scan camera in the background stroboscopically monitors the horizontal position of both pendula as a function of time using narrow reflectors (not seen).}
\label{fig:pendula2}
\end{center}
\end{figure}
To derive the equation of motion, we employ the Lagrange formalism.
For two uncoupled physical pendula the Langrangian reads \cite{Goldstein2001}
\begin{equation}
\mathcal{L}_k
= \frac{1}{2}J_k \dot\varphi_k^2 + J_k\omega_k^2\cos\varphi_k,
\label{Lk}
\end{equation}
where $k=1,2$ labels the two so-far uncoupled pendula and the deflection
angles $\varphi_k$ are zero for vertically hanging pendula described by their moments of inertia $J_k$. In the second term, which is the negative potential energy, we have already made use of the fact that the eigenfrequency reads $\omega_k^2 = gM(l_\text{c})_k/J_k$, with the gravity acceleration $g$, the mass $M$, which is equal for both pendula, and the distance $(l_\text{c})_k$ between the pivot and the center of mass of each pendulum. Note, that up to \eq{shift}, $\varphi_k$ denotes the deflection angles in the lab frame with the vertical direction as reference.

\subsection{Coupling by magnets}

The two pendula are coupled by either one or two pairs of permanent neodymium magnets as sketched in \fig{fig:pendula2}{a} for the case of one pair of magnets. The cubical magnets can be seen in the photograph in \fig{fig:pendula2}{b}. Two large magnets with 28\,mm edge lengths are fixed to the end of each rod and can be rotated by battery driven motors. Two smaller magnets are positioned inside the red cylinders 0.513\,m above. All magnet cubes are magnetized along their horizontal four-fold axes, which at all times remain fixed within the oscillation plane with the excumption of one of the lower magnets optionally rotated around the rod it is attached to.

With the origin set to the pivot of the left pendulum ($k=1$), the magnets have positions
\begin{equation}
\vec r_1 = \begin{pmatrix} l\sin\varphi_1 \\ - l\cos\varphi_1 \\ 0 \end{pmatrix}
,\quad
\vec r_2 = \begin{pmatrix} L+l\sin\varphi_2 \\ - l\cos\varphi_2 \\ 0 \end{pmatrix} ,
\label{vecr}
\end{equation}
where $L$ is the horizontal distance between the pivots of the pendula, and $l$ is the distance between magnet and pivot, which is equal for both pendula. Their magnetic dipole moments of identical magnitude $m$ are oriented as
\begin{equation}
\vec m_1 = m \begin{pmatrix} \cos\varphi_1 \\ \sin\varphi_1\\ 0 \end{pmatrix},
\quad
\vec m_2 = m \begin{pmatrix} \cos\varphi_2\cos(\Omega t) \\ \sin\varphi_2\cos(\Omega t) \\
\sin(\Omega t) \end{pmatrix} ,
\label{vecm}
\end{equation}
valid for large enough $L$ as is the case in all our experiments.

The magnetic dipole-dipole coupling energy is \cite{Jackson1999}
\begin{equation}
\mathcal{U}_{12}
= \frac{\mu_0}{4\pi R^3}
  [\vec m_1\cdot\vec m_2 - 3(\vec m_1\cdot \vec e_R) (\vec m_2\cdot \vec
e_R)] \,,
\end{equation}
where $R\vec e_R = \vec R \equiv \vec r_2-\vec r_1$.
To express this in terms of our dynamical variables $\varphi_k$, we insert the vectors in Eqs.~\eqref{vecr} and \eqref{vecm} and obtain the potential energy of the coupling to read $\mathcal{U}_{12} = \mathcal{V}_{12}\cos(\Omega t)$, where
\begin{equation}
\mathcal{V}_{12}
= \frac{\mu_0 m^2}{4\pi R^3}
  [ \cos(\varphi_1-\varphi_2)
   -3\cos(\varphi_1-\psi)\cos(\varphi_2-\psi)] \,.
\label{intV}
\end{equation}
Length and orientation of $\vec R$ follow from straightforward
geometric considerations and read
\begin{align}
R ={}&
\sqrt{(L+l\sin\varphi_2-l\sin\varphi_1)^2+(l\cos\varphi_2-l\cos\varphi_1)^2} ,
\nonumber
\\
\psi={}&
\arctan\left(\frac{\cos\varphi_1-\cos\varphi_2}
                  {L/l-\sin\varphi_1+\sin \varphi_2}\right) \,,
\label{Rpsi}
\end{align}
where $\psi$ is the angle between the $x$-axis and $\vec R$, cf.\ \fig{fig:pendula2}{}.  In some of our experiments, we use two pairs of magnets, i.e., a lower pair connected to the ends of the rods and an upper pair with a variable horizontal distance shifted upwards along the rods. Consequently, we introduce the respective distances \llow\ (replacing $l$) and \lup\ between pivots and lower versus upper magnets. Hereby, $l_\text{l}-l_\text{u}$ exceeds the horizontal distance $L$ roughly by a factor two, such that we can neglect the interaction between the upper and the lower magnets and consider two separate dipole-dipole couplings. The upper magnets orientations are fixed, such that their magnetizations are perpendicular to the respective rods, remain in the $x$-$y$ plane and attract each other. In contrast, one of the lower magnets is rotated such that
\begin{equation}
\mathcal{U}_{12} = \mathcal{V}_{12}^\text{u} +
\mathcal{V}_{12}^\text{l}\cos(\Omega t).
\end{equation}

The equations of motion follow readily from the Lagrange equation
\cite{Goldstein2001}
\begin{equation}
\frac{d}{dt}\frac{\partial\mathcal{L}}{\partial\dot\varphi_k}
= \frac{\partial\mathcal{L}}{\partial\varphi_k}.
\end{equation}
For the full Lagrangian, $\mathcal{L}=\mathcal{L}_1+\mathcal{L}_2
-\mathcal{U}_{12}$, they become
\begin{equation}
J_k\ddot \varphi_k +J_k\omega_k^2\sin \varphi_k
= - \frac{\partial\mathcal{U}_{12}}{\partial\varphi_k} ,\quad k=1,2.
\label{eom_p1}
\end{equation}
The derivative of the interaction potential is straightforward to calculate but results in a bulky expression, whose explicit form is not needed for our discussion. Equations \eqref{eom_p1} can be integrated numerically to obtain a full solution of the pendula dynamics which considers all nonlinearities.

\subsection{Linearization and quasistatic solution}\label{sec:quasistatic}

The mapping of Newton's equation of motion to a Schr\"odinger equation requires a bilinear Lagrangian, as will become clear in Sec.\ \ref{sec:schroedinger} below. In our experiment, the deflection angles are rather small, $|\varphi_k|\lesssim 1^\circ$, such that already a linearized version of the equation of motion provides a faithful description.  We will perform this linearization by an expansion for the Lagrangian up to second order in $\varphi_k$.  Then the Lagrangian of the uncoupled pendula, Eq.~\eqref{Lk}, takes the bilinear form $\mathcal{L}_k = J_k(\dot\varphi_k^2 - \omega_k^2\varphi^2)/2$, where we have omitted an irrelevant constant term.  The expansion of the interaction term \eqref{intV} requires more effort, because the $\varphi_k$ appear in both the numerator and the denominator.

Closer inspection of the expressions, however, reveals that the $\varphi_k$-dependence of the numerator is much weaker than that of the denominator, given by $R^{-3}(\varphi_k)$, see \eq{Rpsi}.  The difference stems from the numerical pre-factors of the second-order terms of the Taylor expansions of $(1-x)^{-3}$ and $\cos(x)$, which are $12$ and $1/2$, respectively.  Moreover, in $R$ the $\varphi_k$ are weighted by a factor $l/L$, which for our experiment is $\sim 4$.  This leads to the conclusion that the numerator is responsible for only $\sim 1\%$ of the curvature of $\mathcal{U}_{12}$.  Hence, we neglect the higher order terms of the numerator and approximate $\cos(\varphi_1-\varphi_2)-3\cos(\varphi_1-\psi)\cos(\varphi_2-\psi)\simeq-2$. However, we do consider the terms stemming from the expansion of the denominator,
\begin{equation}
\begin{split}
R^{-3}
={}& [L - l\varphi_1+l\varphi_2 +\mathcal{O}(\varphi^4)]^{-3}
\\
\simeq{}& L^{-3} \left( 1 + 3\alpha + 6\alpha^2 + 10\alpha^3\right)
\end{split}
\end{equation}
with $\alpha = (\varphi_1-\varphi_2)l/L$.  Inserting the relevant terms up
to second order into Eq.~\eqref{intV} yields the Langrangian of the
linearized problem
\begin{equation}
\mathcal{L} =
\sum_{k=1,2} \frac{J_k}{2} (\dot\varphi_k^2 - \omega_k^2\varphi_k^2)
+ \tilde F(\varphi_1-\varphi_2)
+ \frac{\tilde G}{2} (\varphi_1-\varphi_2)^2 .
\label{Llin}
\end{equation}
For our two pairs of magnets, where one of the lower magnets is rotating, the energies $\tilde G$ and $\tilde F$ take the form
\begin{equation}
\begin{split}
\tilde G(t) ={}& G_\text{u} + G_\text{l}\cos(\Omega t) ,\\
\tilde F(t) ={}& \frac{L_\text{u}}{4l_\text{u}}G_\text{u} + \frac{L}{4l_\text{l}}G_\text{l}\cos(\Omega t),
\label{FGul}
\end{split}
\end{equation}
where the tilde indicates parameters stemming from second-order Taylor
expansion of the Lagrangian. Equation \eqref{FGul} neglects the
time-dependent equilibrium positions of the pendula, which we will take
into account as a correction below. The coefficients in Eq.~\eqref{FGul},
which quantify the interaction energies
are determined by $\mathcal{U}_{12}$ and read
\begin{equation}
\begin{split}
G_\text{u} ={}& \frac{6\mu_0m_\text{u}^2 l_\text{u}^2}{\pi L_\text{u}^5} ,\\
G_\text{l} ={}& \frac{6\mu_0m_\text{l}^2 l_\text{l}^2}{\pi L^5} ,
\end{split}
\label{FG}
\end{equation}
where $L_\text{u}$ is the distance between the upper magnets at $\varphi_1=\varphi_2=0$. $L_\text{u}$ is varied between experiments and $L_\text{u}=\infty$ corresponds to the case without upper magnets.

To bring $\mathcal{L}$ into a bilinear form, we transform the coordinates such that the linear term $\tilde F(\varphi_1-\varphi_2)$ on the right-hand side of Eq.~\eqref{Llin} vanishes. This can be achieved by introducing coordinates relative to the potential minimum which is located at
\begin{equation}
\begin{split}
\varphi_1^\text{qs}(t)
={}& \frac{J_2\omega_2^2 \tilde F(t)}{J_1J_2\omega_1^2\omega_2^2
     - \tilde G(t)(J_1\omega_1^2+J_2\omega_2^2)},
\\
\varphi_2^\text{qs}(t)
={}& -\frac{J_1\omega_1^2 \tilde F(t)}{J_1J_2\omega_1^2\omega_2^2
     - \tilde G(t)(J_1\omega_1^2+J_2\omega_2^2)} ,
\end{split}
\label{eq:coupling_eq}
\end{equation}
where the superscript ``qs'' refers to the quasistatic solution. The adiabatic equilibrium positions $\varphi_k^\text{qs}$ implicitly depend on the center-of-mass coordinates since $J_k\omega_k^2=Mg(l_\text{c})_k$.

Our driving frequency is much smaller than the resonance frequencies of the two pendula, $\Omega/\omega_0\lesssim0.01$, where $\omega_0=(\omega_1+\omega_2)/2$. Hence, within an adiabatic approximation, \eq{eq:coupling_eq} describes the steady-state solution of the coupled pendula driven by the rotation of one of the magnets. Henceforth, we use $\varphi_k^\text{qs}(t)$ as reference point, which is achieved by the transformation
\begin{equation}
\varphi_k \to \varphi^\text{lab}_k = \varphi_k + \varphi_k^\text{qs}(t)\,.
\label{shift}
\end{equation}
Moreover, we neglect within an adiabatic approximation the time derivatives of $\varphi_k^\text{qs}(t)$ which removes the linear term in Eq.~\eqref{Llin} and results in the desired intended bilinear Lagrangian.

By applying the transformation \eqref{shift} the corresponding equation of motion becomes not only of first order and linear, but also homogeneous. Accordingly, it assumes the form of a Schr\"odinger equation. However, the coordinates $\varphi_k$ are no longer in the lab frame. Instead, they are the deflection angles with respect to the time-dependent quasistatic solution in Eq.~\eqref{eq:coupling_eq}.

\subsection{Dynamic potential curvature}
\label{sec:3rd-order}

The price for separating $\varphi_k$ from the time dependence of the
equilibrium positions of the pendula is that for the new coordinates the
interaction becomes time dependent even for constant $\varphi_1$ and
$\varphi_2$. In a hand-waving picture, while one magnet is rotating the
equilibrium distance between the magnets is smaller whenever the
interaction is attractive as compared to the case of repulsive interaction.
As a result, the interaction itself is accordingly modulated for any values
of $\varphi_1$ and $\varphi_2$. We will now quantify this correction.

So far, we have linearized the equation of motion by performing a Taylor
expansion of the interaction potential at the pivot distance $L$.  However,
after the transformation \eqref{shift}, the potential curvature at
$\varphi_1=\varphi_2$ is no longer given by the energy $\tilde G(t)$ in
Eq.~\eqref{FGul}, but by the corresponding Taylor coefficient evaluated at
$L-\lup\delta\varphi^\text{qs}(t)$ for $G_\text{u}$ and
$L-\llow\delta\varphi^\text{qs}(t)$ for $G_\text{l}$. This can be captured
by the approximate correction
\begin{equation}
\begin{split}
\tilde G(t) \to G(t)
={} &
G_\text{u} \Big[1-\frac{l_\text{u}}{L_\text{u}}\delta\varphi^\text{qs}(t)\Big]^{-5}
\\ & + 
G_\text{l} \cos(\Omega t)
\Big[1-\frac{l_\text{l}}{L}\delta\varphi^\text{qs}(t)\Big]^{-5} .
\end{split}
\label{G'}
\end{equation}
Clearly, the interaction term acquires a further time dependence via
$\delta\varphi^\text{qs}(t)\equiv\varphi_1^\text{qs}(t)-\varphi_2^\text{qs}(t)$
in addition to the modulation expressed in Eq.~\eqref{FGul}.  The explicit
dependence of $G(t)$ on $\delta\varphi^\text{qs}(t)$ results in an
asymmetry between the times of attractive versus repulsive magnetic force
even without upper magnets.

To see the consequences of this correction, we restrict ourselves to weak
asymmetries such that the pendulum frequencies and moments of inertia can
be replaced by their average values $\omega_0$ and
$J_0 = (J_1+J_2)/2$.  Then we arrive at the approximation
\begin{equation}
\varphi_1^\text{qs}(t) = \frac{\tilde F(t)}{\omega_0^2 J_0-2\tilde G(t)} = -\varphi_2^\text{qs}(t) ,
\label{qs_approx}
\end{equation}
which allows the numerical evaluation of the effective potential curvatures using Eq.~\eqref{G'}.

To make analytical progress, we restrict ourselves to the case in which only the rotating lower magnets are present, while $G_\text{u} = 0 = F_\text{u}$.  Our aim is to show that, in consistency with the experimental observation, nevertheless the effective potential curvature has a non-vanishing mean value.  In doing so, we keep only corrections to lowest order in $G(t)$ and, hence, in $\delta\varphi^\text{qs}(t)$ such that the effective potential curvature becomes
\begin{equation}
\begin{split}
G(t) & \simeq
G_\text{l}\cos(\Omega t) \Big[1 + \frac{5}{2}\frac{G_\text{l}\cos(\Omega t)}{\omega_0^2J_0}\Big]
\\
& = G_\text{l}\cos(\Omega t)
+ \frac{5 G_\text{l}^2}{4\omega_0^2J_0} [1+\cos(2\Omega t)] .
\end{split}
\end{equation}
The first term describes the modulation of the interaction by the rotating magnet in accordance to \eq{FGul}. The second constant term describes the average increase of the interaction related to the fact that the dipole-dipole interaction is more enhanced during attraction than reduced during repulsion. The third term describes a second harmonic modulation in $\Omega$.  The latter merely distorts the shape of the driving and can be neglected on the level of agreement we are aiming at.

Let us emphasize that the derivation of the correction \eqref{G'} relies on
the assumption that the nonlinear $\varphi_k$ dependence of the interaction
$\mathcal{U}_{12}$ can be captured by a time-dependent quadratic term.
While this reasoning naturally has limitations, it clearly reveals that the varying
potential curvature causes an effective constant interaction term, even in
the absence of the upper magnets. It explains the observed dependences
$\varepsilon_0(L)$ and $A(L)$ without upper magnets, cf.\ the red line in
\fig{fig:LZSM}b or \fig{fig:AEeff}{} below.

\section{Schr\"odinger-like Equation}\label{sec:schroedinger}
\label{app:schrodinger}

Next, we bring the linearized equation of motion of our system to the form of a
Schr\"odinger equation that describes the time-dependent amplitude of the
fast oscillations of each pendulum with approximately the average frequency
$\omega_0 = (\omega_1+\omega_2)/2$.  The pedestrians way
\cite{SeitnerPRB16,ShoreAJP09,NovotnyAJP10} starts from the linearized
classical equation of motion
which is of second order. To obtain a differential equation of first order,
one employs for the deflection angles the complex ansatz $\varphi_k = \Re
e^{-i\omega_0t}\Psi_k + \text{c.c.}$, $k=1,2$, where $\Psi_k$ is a slowly varying
amplitude. Within a rotating-wave approximation, one then neglects the
second-order derivatives of $\Psi_k$ and all terms that oscillate with the
angular frequency $\omega_0$.

Here we pursue a more elegant alternative by performing these steps within
the Lagrange formalism. Accordingly, our goal is to transform
Eq.~\eqref{Llin} such that it assumes the form of the Lagrangian of the
Schr\"odinger equation $i\hbar\partial_t\Psi = H\Psi$, which reads
\begin{equation}
\mathcal{L}_\text{Schr}
= i\hbar\Psi^\dagger \partial_t\Psi - \Psi^\dagger H \Psi ,
\label{Lschr}
\end{equation}
where $\Psi$ and  $H$ denote the probability amplitudes and the Hamiltonian in vector-matrix notation. Alternatively, one may use the symmetrized form of the Lagrangian, $\mathcal{L}_\text{Schr} = i\hbar(\Psi^\dagger\partial_t\Psi-\Psi\partial_t\Psi^\dagger)/2 - \Psi^\dagger H \Psi$, which differs from \eq{Lschr} by an irrelevant total time derivative. Its relation to the Schr\"odinger equation follows readily from the Langrange equation
\begin{equation}
\frac{d}{dt} \frac{\partial\mathcal{L}}{\partial\dot\Psi_k^*}
= \frac{\partial\mathcal{L}}{\partial\Psi_k^*} .
\end{equation}
The fact that $\mathcal{L}_\text{Schr}$ is bilinear, makes it obvious that it was indispensable to remove the term of \eq{Llin} linear in $\varphi_k$ via the transformation \eqref{shift} and to avoid explicit terms of higher order. 

Let us stress that the resulting equation still describes classical mechanics and, despite its form, does not constitute a quantization.  In particular, the quantum mechanical energy-frequency relation given by Planck's constant does not hold. Technically, this is not a problem as long as we use a Schr\"odinger equation in the dimensions of frequency. If we assume $\hbar=1$, it is identical to the usual quantum version of the Schr\"odinger equation in units of frequency.

\subsection{Hamiltonian and conservation law}

Like in the standard approach, we assume that the average pendulum frequency $\omega_0$ is much larger than all other frequency scales, which suggests an ansatz with a rapid oscillation and a slowly varying amplitude. To be specific, we define
\begin{equation}
\varphi_k(t) = \Psi_k(t) e^{-i\omega_0 t} + \text{c.c.} ,
\label{app:ansatz}
\end{equation}
where $\Psi_k$ is generally complex.  While inserting this ansatz into Eq.~\eqref{Llin}, we keep only terms that are at least of order $\omega_0$. (The ansatz is constructed such that terms $\propto\omega_0^2$ cancel each other.)  Moreover, we neglect within a rotating-wave approximation all terms that contain the phase factor $e^{\pm i\omega_0t}$.

In the resulting expression, the remaining part of the first term of
$\mathcal{L}$ in Eq.~\eqref{Llin} becomes
$2i\omega_0J_k\Psi_k\partial_t\Psi_k$.  It is still not of the desired
form, because its pre-factor still depends via the moment of inertia on the
mode index $k$. To reach the form of a Schr\"odinger equation, one needs
pre-factors independent of the mode index $k$, e.g., by re-scaling the amplitude with a factor $J_k^{-1/2}$.  This however
would no longer allow the intuitive interpretation of the $\Psi_\pm$
defined below as in-phase and out-of-phase modes. To nevertheless get rid of the $k$-dependence of the
pre-factors, we replace the $J_k$ by their average $J_0 = (J_1+J_2)/2$.
This is a reasonable approximation for most of our experiments, as they
fulfill $|J_1-J_2| \ll (J_1+J_2)$.  Finally, for convenience we divide by
$2\omega_0J_0$ (which has no consequence for the equations of motion) to
obtain a Lagrangian with dimension frequency
\begin{equation}
\label{Ltls}
\mathcal{L}
= i\sum_k \Psi_k^*\dot\Psi_k
     -\frac{\Delta}{2}(|\Psi_1|^2-|\Psi_2|^2) + \frac{\varepsilon}{2} |\Psi_1-\Psi_2|^2 ,
\end{equation}
where we have introduced the frequency difference
\begin{equation}
\label{tls.delta}
\Delta = \omega_1-\omega_2
\end{equation}
and the time dependent coupling (in units of frequency)
\begin{equation}
\label{tls.epsilon}
\varepsilon(t) = \frac{G(t)}{\omega_0 J_0} .
\end{equation}
Comparison with Eq.~\eqref{Lschr} demonstrates that the Lagrangian in
Eq.~\eqref{Ltls} corresponds to a Schr\"odinger equation of a
two-level-system (TLS) with the Hamiltonian
\begin{equation}\label{H_local}
H = \frac{1}{2}\begin{pmatrix}
    \Delta -\varepsilon(t) & \varepsilon(t) \\
    \varepsilon(t) & -\Delta -\varepsilon(t) \end{pmatrix}
\end{equation}
in units of frequency with $\hbar=1$.

The $U(1)$ symmetry of the Lagrangian \eqref{Ltls} together with Noether's
theorem immediately provides the conservation of $N \equiv |\Psi_+|^2 +
|\Psi_-|^2$.  Then the energy $\mathcal{E} = \mathcal{T}+\mathcal{U}$ of
the pendula obeys the proportionality
\begin{equation}
\mathcal{E}
= \omega_0^2 N + 2\omega_0 \langle H\rangle .
\end{equation}
It is dominated by the constant term $\omega_0^2 N$, while the expectation
value $\langle H\rangle$ represents a small, generally time-dependent
correction to the energy of the driven coupled pendula.  Because of the
driving, normalization of the wave function corresponds to a merely approximate energy conservation of the pendula motion.

\subsection{Basis transformation}\label{sec:basistransformation}

The preferential basis in most works on quantum dots is the one formed by
localized states.  Then ac voltages applied via plunger gates appear in the
Hamiltonian as time-dependent diagonal elements.  To establish a closer
connection to these systems, we introduce the in-phase and out-of-phase
mode $\varphi_\pm = \varphi_1\pm\varphi_2$ and the corresponding envelopes
$\Psi\pm = \Psi_1\pm\Psi_2$.  This corresponds to a unitary transformation
with $S = (\sigma_x+\sigma_z)/\sqrt{2}$, which formally interchanges the
Pauli matrices $\sigma_x$ and $\sigma_z$.  With $H \to SHS^\dagger$
our Schr\"odinger equation takes the form
\begin{equation}
\label{Htls}
i\frac{d}{dt}\begin{pmatrix} \Psi_+ \\ \Psi_- \end{pmatrix}
=
\begin{pmatrix}
   0 & \Delta/2 \\
   \Delta/2 & -\varepsilon(t)
\end{pmatrix} \begin{pmatrix} \Psi_+ \\ \Psi_- \end{pmatrix} ,
\end{equation}
which represents the basis of the analogy between the pendula and
the quantum mechanical two-level system.

\subsection{Duality and limitations}\label{sec:duality}

\begin{table}
\caption{Correspondence between a qubit described by the Schr\"odinger equation and the classical coupled pendula after the rotating-wave approximation, which alters the EOM independent of the carrier frequency $\omega_0 = (\omega_L+\omega_R)/2$. Each line contains two corresponding quantities.}
\label{table:params}
\begin{ruledtabular}
\begin{tabularx}{\columnwidth}{lr}
two-level system & coupled pendula
\\\hline\\[-1.5ex]
eigenstates & normal modes
\\
tunnel oscillations & beating
\\[1.5ex] 
tunnel coupling $\Delta$ & frequency diff. $\Delta = \omega_1-\omega_2$
\\
energy detuning $\varepsilon(t)$ & interaction $\varepsilon(t)$
\\[1.5ex] 
localized states & in-phase/out-of-phase mode
\\
delocalized states & left/right pendulum mode
\\[1.5ex] 
amplitude of wavefunctions & amplitude of pendula
\\
occupation probability & occupation $\propto$ energy 
\end{tabularx}
\end{ruledtabular}
\end{table}

We have chosen the diabatic basis spanned by the in-phase and out-of-phase modes.  The corresponding delocalized basis for the Hamiltonian in Eq.~\eqref{Htls} corresponds to a localized basis usually employed for LZSM physics with a qubit, e.g., the one defined by the localized electron states in a double quantum dot \cite{ForsterPRL14}. In Table~\ref{table:params} we present the cross relation between the localized and delocalized bases corresponding to mapping our coupled pendula to a double dot qubit.

An interesting point arises from the fact that we modulate the coupling between the pendula while in qubits usually the detuning between localized states is modulated. The consequence is a duality between the terms ``detuning'' and ``coupling'' in Table~\ref{table:params}: The coupling of the pendula by the magnetic interaction corresponds to the detuning of the quantum dot levels, while the detuning of the pendulum frequencies corresponds to the tunnel coupling.

The mapping to a Schr\"odinger equation relies on the slowly-varying envelope approximation which requires $\Delta, \varepsilon\ll\omega_0$. While usually the case, for our smallest pivot distance, $L=208\,$nm, this is fulfilled only marginally as we find $|\varepsilon|$-values up to $\SI{1}{\s}^{-1}$ while $\omega_0\simeq3.3\,\text{s}^{-1}$.  Indeed, such large $|\varepsilon|$-values already significantly reduce the potential curvatures in some direction such that with increasing coupling, one eigenmode eventually becomes unstable.  This happens when for $\varepsilon(t)>\omega_0$ the adiabatic eigenfrequencies of the linearized Newton equation become imaginary during some time intervals, see \fig{fig:lzsmfan}{} below and its discussion in \sect{sec:LZSM}.

Finally, we comment on the approximation by which we replaced the moments of interia $J_1$ and $J_2$ by their average, i.e., we neglected terms propotional to $\delta J = J_1-J_2$.  If we had kept this term and employed the ansatz \eqref{app:ansatz}, our Hamiltonian would have acquired a non Hermitian contribution.  Its size might be reduced by a more sophisticated ansatz, but then the modes would depend on the $J_k$ and, thus, on the parameter sets. For our LZSM-interferometry experiments, $\delta J\ll J_{1,2}$, such that practical corrections are minor.  In the case of our Rabi experiments $\delta J\le 0.2 J_{1,2}$, which is sufficient for the weak couplings considered there.

\section{Driven qubit dynamics}
\label{app:drivenqubit}

So far, we have shown that within the range of validity of the
rotating-wave approximation, the oscillations of our coupled pendula have
an envelope which obeys the Schr\"odinger equation of the TLS.  Next we
turn to the particular case of a periodically time-dependent interaction
which maps to a time-dependent TLS detuning $\varepsilon(t) = \varepsilon_0
+ A\cos(\Omega t)$, while higher harmonics do not play a relevant role. In
this case the Hamiltonian in \eq{Htls} takes the form
\begin{equation}
H(t) = \frac{\Delta}{2}\sigma_x +\frac{\varepsilon(t)}{2}\sigma_z ,
\label{app:H(t)}
\end{equation}
where we have added an irrelevant term proportional to the unit matrix
such that $H(t)$ becomes traceless.
In the following, we review the typical quantum phenomena observed with this Hamiltonian.

\subsection{Rabi oscillations}\label{sec:Rabi}

The most prominent textbook example is the Rabi problem found for resonant
driving with frequency $\Omega \simeq \Delta$, small amplitude
$A\ll\Omega,\Delta$ and zero detuning, $\varepsilon_0=0$ \cite{CohenTannoudji1992}.
In this limit, it is convenient to work in the eigenbasis of the
undriven (here uncoupled) system, henceforth denoted by a tilde, which corresponds to the
basis of individual pendulum modes.  Then the Hamiltonian
\eqref{app:H(t)} reads
\begin{equation}
\widetilde H(t) = \frac{\Delta}{2}\sigma_z
+ \frac{A}{2} \cos(\Omega t) \sigma_x .
\end{equation}
In the quantum optical context, this model describes atomic transitions induced by irraditation with a laser that couples to the dipole moment of the atom.

It is now convenient to transform the Hamiltonian via the unitary $S =
e^{-i\sigma_z\Omega t/2}$ to a rotating frame, $\widetilde H \to S^\dagger
\widetilde H S -iS^\dagger\dot S$.  Sufficiently close to resonance and for
small driving amplitude, $|\Delta-\Omega|\ll\Omega$ and $A\ll\Omega$, one
can neglect within a rotating-wave approximation rapidly oscillating terms
to find the time-independent two-level Hamiltonian
\begin{equation}
\widetilde H_\text{eff}
= \frac{1}{2} \begin{pmatrix} \Delta-\Omega & \omr \\
\omr & -\Delta+\Omega \end{pmatrix}
\end{equation}
with the Rabi frequency
\begin{equation}\label{eq:Rabi}
\omr = \frac{A}{2}
= \frac{3\mu_0\mlow^2 l_\text{l}^2}{\pi\omega_0 J_0 L^5}
\simeq\frac{3\mu_0\mlow^2 l_\text{l}^2}{\pi M g^2}\,\frac{\omega_0^3}{L^5}\,.
\end{equation}
Hence, the occupation probabilities oscillate with the frequency
\begin{equation}\label{eq:effRabi}
\Omega_\text{eff} = \sqrt{(\Delta-\Omega)^2 + \omr^2} \,.
\end{equation}
The latter approximation assumes $J_0=M\ell^2$
and $\omega_0^2 = g/\ell$ (mathematical pendulum), such that we can
eliminate $J_0\simeq Mg/\omega_0^4$. For resonant driving $\Omegam=\Delta$,
the resulting dynamics consists of Rabi oscillations between the ground
state and the excited state with frequency \Omegar\ with full probability
transfer, i.e., visibility $\nu=1$. Thereby, the lower frequency eigenmode
of the coupled pendula corresponds to the ground state of the quantum
mechanical TLS. Note, that in our Rabi-oscillation experiments no upper
magnets are present and, moreover, $A\ll\Delta$, such that also the
effective static interaction derived in Sec.~\ref{sec:3rd-order} is
negligible, $\varepsilon_0\simeq0$.

\begin{figure}
\centerline{\includegraphics{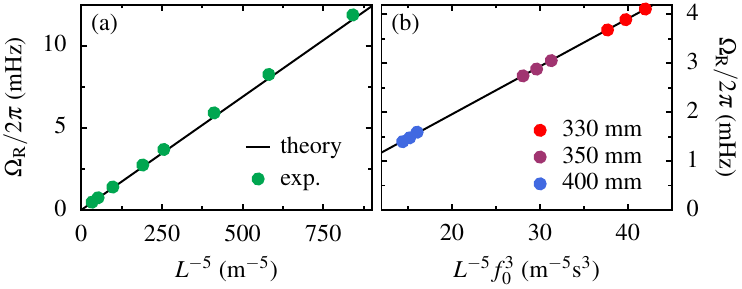}}
\caption{Probing the magnetic dipole-dipole interaction on Rabi oscillations (no upper magnets are used). The symbols are measured Rabi frequencies while the straight lines are calculated with \eq{eq:Rabi}. The propotionality $\omr\propto L^{-5}$ is a consequence of the interaction energy $G_\text{l}$ in \eq{FG}. Model lines are calculated with material parameters listed in Tables \ref{table:params_individual} and \ref{table:params_coupled}, there is no free fit parameter.
(a) Rabi frequency as a function of $L^{-5}$ for varying pivot distance $L$;
(b) Rabi frequency as a function of  $L^{-5}f_0^3$ for three different $L$ (as labeled) and varying $f_0 = \omega_0/2\pi$, which is the average pendulum eigenfrequency.
\label{fig:Rabi-L}
}
\end{figure}

We used \eq{eq:effRabi} for the model data shown in \fig{fig:rabi}c plotting $\Omega_\text{eff}(\Delta)$. In \fig{fig:Rabi-L}{}
we additionally compare \eq{eq:Rabi} with the measured Rabi frequency while we varied the pivot distance $L$ and the average eigenfrequency of both pendula. For the model curves we used the setup parameters listed in Tables \ref{table:params_individual} and \ref{table:params_coupled}. The good agreement between model data and theory in all three cases confirms the validity of our approximations within the regime of small couplings realized in our Rabi measurements.

\subsection{Single Landau-Zener passage}

The adiabatic eigenenergies of the Hamiltonian \eqref{app:H(t)} exhibit an avoided crossing at $\varepsilon=0$.  For large driving amplitudes $A$, a single traverse of such an avoided crossing is a standard problem in time-dependent quantum dynamics.  In an idealized model, one linearizes the time-dependent detuning to obtain
$\varepsilon(t) = vt$ with the sweep velocity 
\begin{equation}
v = \pm\Omega\sqrt{A^2-\varepsilon_0^2}
\end{equation}
at the crossing. Then the Hamiltonian can be approximated by an idealized version with linearized time-dependence,
\begin{equation}
H_\mathrm{LZ}(t) =
\frac{1}{2}\begin{pmatrix} vt & \Delta \\ \Delta & -vt \end{pmatrix}
=\frac{1}{2}\left(vt\sigma_z + \Delta\sigma_x\right) .
\label{eq:lzelin}
\end{equation}
Since the dynamics takes place close to the crossing, one may extend the
time range to infinity, which allows an analytic solution.  In 1932,
Landau, Zener, St\"uckelberg, and Majorana \cite{LandauPZS32a,
ZenerPRSLA32, StueckelbergHPA32, MajoranaNC32} in four independent works
found that if the system at time $t=-\infty$ is in its adiabatic ground
state, it will in the limit $t\to\infty$ occupy the excited adiabatic state
with the so-called Landau-Zener probabilty
\begin{equation}\label{pLZ}
P_{\rm LZ}=\exp\left(-\frac{\pi \Delta^2}{2 |v|}\right).
\end{equation}
In particular for $|v|\ll\Delta^2$, the system adiabatically follows the initialized ground state, while for $|v|\gg\Delta^2$, it non-adiabatically switches to the excited state.

\subsection{LZSM interference}\label{sec:LZSM}

\begin{figure}
\centerline{\includegraphics{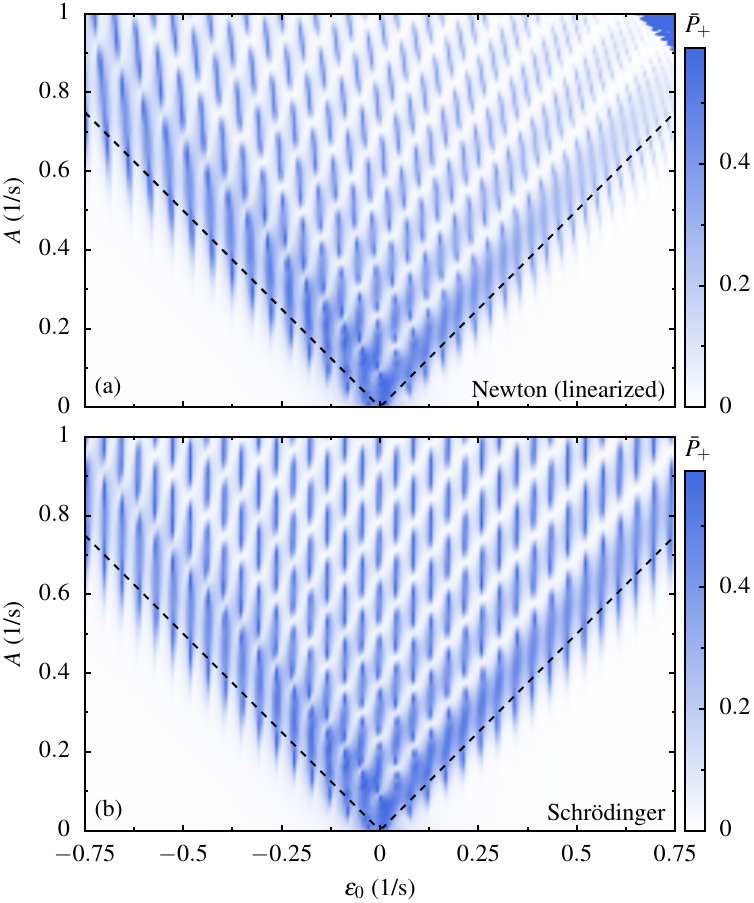}}
\caption{LZSM fan diagram presenting $\overline P_+(\varepsilon_0,A)$ computed (a) with the linearized Newton's equation and (b) with the slowly varying envelope approximation which has the form of a Schr\"odinger equation.  The dashed lines indicate $A=|\varepsilon_0|$, below which the avoided crossing at $\varepsilon(t)=0$ is never reached during the modulation.
}
\label{fig:lzsmfan}
\end{figure}

A more recent topic in ac-driven quantum dynamics is the behavior of a quantum system that is repeatedly swept through an avoided crossing.  Then each crossing acts as beam splitter in energy space, such that one observes interference patterns as a function of the average detuning $\varepsilon_0$ and the driving amplitude $A$ \cite{ShevchenkoPR10}.  Since the crossing condition $\varepsilon(t)=0$ can only be fulfilled for sufficiently large amplitude, $A\gtrsim|\varepsilon_0|$, a nontrivial interference pattern is found in the triangle that meets this condition during the modulation, cf.\ \fig{fig:lzsmfan}{}.
We calculated the interference pattern displayed in \fig{fig:lzsmfan}a based on the linearized Newton's equation, while for the one in \fig{fig:lzsmfan}b we used the Schr\"odinger equation \eqref{Htls} which results from a slowly-varying envelope approximation. The LZSM fan diagram computed with Newton's equation contains a clearly visible distortion for large values of $\varepsilon_0$ and $A$, which is missing in the corresponding calculation using the Schr\"odinger equation. These distortions are related with the reduced potential curvatures discussed in \sect{sec:duality} above.  Moreover, in the upper-right angle of \fig{fig:lzsmfan}a, we observe a region of saturate amplitude pointing to an instability. It emerges when the coupling exceeds the pedulum frequencies, $\varepsilon(t)\gtrsim\omega_0$ at least at some instances of time, such that one eigenvalue of the linearized Newton equation (Eq.~(1) of the main text) becomes imaginary. Both, the distortion and the instability pose a limit to a one-to-one comparison between coupled pendula and a quantum mechanical two-level-system.

We demonstrate a classical analog to a qubit within this limitation [demonstrated by the differences between \fig{fig:lzsmfan}{a} and \fig{fig:lzsmfan}{b}]. Beyond, we provide clear evidence for LZSM interference in a macroscopic classical system. In quantum systems, LZSM interference patterns of this kind have been observed for the non-equilibrium population of superconducting qubit \cite{OliverS05, SillanpaaPRL06, WilsonPRL07, IzmalkovPRL08, BernsNL08, LiNC13}, the current through double quantum dots \cite{StehlikPRB12, DupontFerrierPRL13, ForsterPRL14, ForsterPRB15b}, and the response of a cavity coupled to a double quantum dot \cite{KoskiPRL18, MiPRB18, ChenPRB21}.

\section{Data analysis}
\label{app:dataanalysis}

\subsection{Processing of experimental data}\label{sec:processing}

The dynamics of the pendula covers a broad spectrum of time scales, dominated by the fastest one, the  oscillation of the pendula near their resonances with $\omega_0/2\pi\sim0.5\,$Hz. The modulation of the interaction with a frequency $\Omega/2\pi\sim5\,$mHz defines a much slower time scale. A central quantity in our experiments is the modulated interaction quantified by $(G_\text{u}+G_\text{l})/2\pi\omega_0J_0$. It defines the time scale of an envelope of the pendula oscillations, which is slow compared to $\omega_0$. In case of the Rabi experiments it is the smallest time scale, while in our Landau-Zener experiments the modulation is the smallest time scale. In both cases, the resulting beating dynamics can be described by a Schr\"odinger-like equation and, consequently, can be compared with the dynamics of a qubit.

\begin{figure}
\centerline{\includegraphics{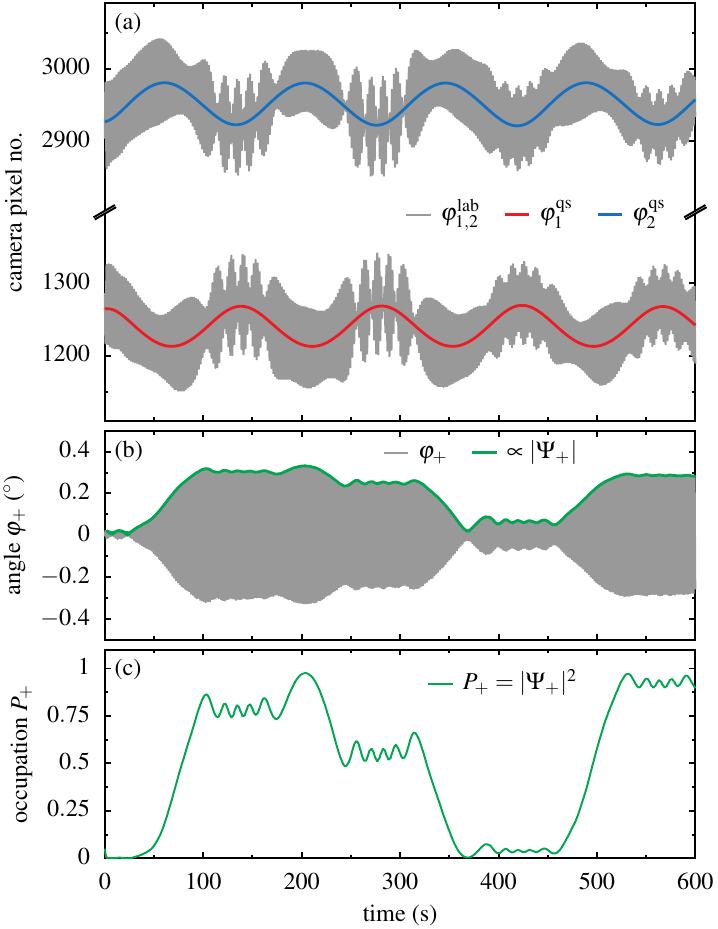}}
\caption{Determining the population dynamics from raw data.
(a) Measured horizontal positions (corresponding camera pixels) as a function of time for both pendula during a LZSM experiment (gray). Apart from an offset, they are proportional to the pendulum angles used as labels in the figure. Individual oscillations are not resolved.  The quasistatic solutions (colored solid lines) are determined by applying a numerical low-pass filter to the raw data.
(b) In-phase mode $\varphi_+(t)=[\varphi_1(t)+\varphi_2(t)]/2$ (gray) as determined from the raw data in panel a after subtracting the $\varphi_k^\text{qs}(t)$. The green line is the envelope $|\Psi_+|$.
(c) Occupation probability $P_+=|\Psi_+|^2$.}
\label{fig:dataanalysis}
\end{figure}
In \fig{fig:dataanalysis}a we present the horizontal positions of both pendula as a function of time directly measured with the line scan camera visible in the background of \fig{fig:pendula2}{}. From these, it is straightforward to calculate the deflection angles, $\varphi^\text{lab}_k$, via a geometric relation. The approximately 300 pendula oscillations with angular frequencies close to $\omega_0$ are not resolved in \fig{fig:dataanalysis}a, which spans a time duration of 10 minutes. We decompose these motions into the rapid oscillations $\varphi_k$, centered at some unknown value $\xi_k$, which by and large is the adiabatic equilibrium position $\varphi_k^\text{qs}$, i.e., we set
\begin{equation}
\varphi^\text{lab}_k = \varphi_k+\xi_k
\label{phiraw}
\end{equation}
with $k=1,2$. The $\varphi_k^\text{qs}(t)$ are shown as colored solid lines in \fig{fig:dataanalysis}a. Owing to the driving, they oscillate with the slow modulation frequency $\Omega$.

In a first step of processing the data, we determine $\xi_k$ using the fact that its dynamics is much slower than the bare pendula oscillations.  On the average, over a few periods of the pendula oscillations, the $\varphi_k(t)$ vanish, such that $\overline{\varphi^\text{lab}_k}= \xi_k(t)$. Hence, subtraction of this time average from the measured deflection angles $\varphi^\text{lab}_k(t)$ yields the rapid oscillations
\begin{equation}
\varphi_k(t)=\varphi^\text{lab}_k(t)-\overline{\varphi^\text{lab}_k}\,.
\label{phik}
\end{equation}
Provided the well separated timescales, a convenient way for performing the time average is a convolution of $\varphi^\text{lab}_k$ with a Gaussian of width $\sim\,10/\omega_0$. Note, that the precise width is practically irrelevant.

For our Rabi experiments we then continue analyzing $\varphi_k$ while for
experiments in the Landau-Zener regime we consider
$\varphi_\pm=(\varphi_1\pm\varphi_2)/2$. In \fig{fig:dataanalysis}b we plot
$\varphi_+(t)$ determined from the raw data shown in
\fig{fig:dataanalysis}a. As the fast oscillations near $\omega_0$ are not
resolved in this plot, the data appear as a gray region with modulated height.

The envelope of this modulation, which is a consequence of the driving, is $|\Psi_+|$. To actually determine the envelope dynamics of $\varphi_k(t)$, we square both sides of Eq.~\eqref{app:ansatz} such that the right-hand side becomes $2|\Psi_k|^2$ plus two terms that oscillate with angular frequency $2\omega_0$.  These rapidly oscillating terms can be removed by convolution with a Gaussian as described above, which provides
\begin{equation}
2 |\Psi_k(t)|^2 = \overline{|\varphi_k|^2},
\label{modPsi}
\end{equation}
where $k=1,2,+,-$. Notice that with this procedure, we cannot obtain the phases of $\Psi_{k}$, hence cannot determine $|\Psi_\pm|$ from $|\Psi_{1,2}|$ or vice versa. Instead, both must be computed from $\varphi_k$ with Eq.~\eqref{modPsi}.

To emphasize the correspondence to the probability amplitude of a qubit, in \fig{fig:dataanalysis}c we show the ``occupation probability'' $P_+(t) = |\Psi_+(t)|^2$ normalized, such that $P_1+P_2 = P_++P_- = 1$.
It visualizes the energy transfer between the
diabatic modes $\varphi_\pm$ for four subsequent passages through their
avoided crossing, while we modulated the magnetic coupling between the
pendula. The first pronounced step can be interpreted as a standard
Landau-Zener transition, while the three subsequent steps are heavily
influenced by the phase development between the pendula oscillations. In
between the pronounced steps occur oscillations on a faster time scale,
also visible in the individual pendula oscillations in
\fig{fig:dataanalysis}a. The time scale of these beats corresponds to the
(modulated) coupling between the pendula. While the coupling clearly
exceeds the modulation frequency in our LZSM experiments, a look at \fig{fig:rabi}{} clarifies, that in the case of Rabi experiments the
modulation frequency exceeds the coupling.

\subsection{Effective two-level-system parameters}\label{sec:TLS-parameters}

To obtain numerical data from the Schr\"odinger equation \eqref{Htls}, we need to know the effective parameters of the driven TLS, namely $\Delta$, $\varepsilon_0$ and $A$.  While the frequency detuning follows readily from the oscillation frequencies of the uncoupled pendula, $\Delta=\omega_1-\omega_2$, the interaction parameters $A$ and $\varepsilon_0$ require more effort. A straightforward but tedious strategy is based on the Newton equation of the setup with all relevant quantities such as the center of mass, the moments of inertia, the distance between the pivots and the magnetic moments. The effective TLS parameters can then be approximated via a Taylor expansion at the equilibrium position.

An additional difficulty is related with our choice of using magnetic dipoles to generate an interaction between the pendula. On the one hand, it allows us to conveniently modulate the coupling, on the other hand the dipole-dipole interaction gives rise to higher-order terms in the expansion of the interaction potential, discussed above in Sec.~\ref{sec:3rd-order}.  For example, in the absence of the upper magnets, Eq.~\eqref{tls.epsilon} predicts for the TLS the driving amplitude $\tilde A = G_\text{l}/\omega_0J_0$, where the tilde indicates that we ignore the time-dependence of the potential curvature as in \eq{FGul}.  Moreover, together with Eq.~\eqref{G'}, it implies that the static TLS detuning $\varepsilon_0$ (i.e.\ the time-averaged coupling of the pendula) stemming from the third-order term of the potential relates to the TLS driving amplitude according to
\begin{equation}
\varepsilon_0 = \frac{5\tilde A^2}{4\omega_0}\,, \qquad A = \tilde A\,.
\label{AEanalyt}
\end{equation}
While this expression provides the correct order of magnitude, comparison with our experimental data reveals that it overestimates $\varepsilon_0$ substantially (by up to 40\%).

To improve our prediction, we refine the above approach by directly
evaluating the effective interation $G$ in Eq.~\eqref{G'} together with the
quasi static position \eqref{qs_approx} without Taylor expansion of the
denominator.  For convenience, as in the Hamiltonian \eqref{app:H(t)}, we
express the interaction energy in terms of the uncorrected TLS parameter
$\tilde\varepsilon(t)=\tilde G(t)/\omega_0J_0$, where we approximated
the eigenfrequencies and moments of inertia of the pendula by their
averages $\omega_0$ and $J_0$, respectively.  Then the quasi static
position reads
\begin{equation}
\delta\varphi^\text{qs} =
\frac{2\tilde F(t)/\omega_0J_0}{\omega_0-2\tilde\varepsilon(t)},
\end{equation}
while \eq{G'}, which expresses the effective $G(t)$ in terms of the
uncorrected $\tilde G(t)$, can be replaced by an improved relation between
the effective interaction $\varepsilon(t)$ and the uncorrected interaction
$\tilde\varepsilon(t)$,
\begin{equation}\label{epsilonbar}
\varepsilon(t) = \tilde\varepsilon(t)\Big(1-\frac{l_\text{l}}{L}\delta\varphi^\text{qs}\Big)^{-5}.
\end{equation}
If we again approximate the time dependence by its symmetrized form, $\varepsilon(t)=\varepsilon_0+A\cos(\Omega t)$, we obtain the effective parameters given by the first two Fourier coefficients of
$\varepsilon(t)$, namely
\begin{align}
\varepsilon_0 ={}& \int_0^T\frac{dt}{T} \varepsilon(t) ,
\\
A = {}& 2\int_0^T\frac{dt}{T} \varepsilon(t) \cos(\Omega t) .
\end{align}

While the prediction of $A$ and $\varepsilon_0$ from \eq{epsilonbar} surpasses that of \eq{AEanalyt}, it still uses an expansion which looses accuracy with increasing coupling strength between the pendula. To circumvent this problem, below we follow an alternative approach, where we determine $A$ and $\varepsilon_0$ from the measured dynamics. This experimental approach is still based on the conjecture that the observed dynamics can be described by the Schr\"odinger equation \eqref{Htls} with $\varepsilon(t) = \varepsilon_0 + A\cos(\Omega t)$. In the following, we describe three complementary methods by which this task may be performed and explain, why the third method provides the most accurate results.

\subsubsection{Fourier analysis}

A rather direct method to determine $A$ and $\varepsilon_0$ from measurements is based on the Fourier spectra of the diabatic modes $\varphi_\pm$ or, equivalently, $\Psi_\pm$. It works best, if $\varepsilon(t)>\Delta$ most of the time such as in LZSM experiments. Then the Hamiltonian is dominated by the interaction $\varepsilon(t)$, while the detuning $\Delta$ can be neglected. Under such conditions, the mode $\Psi_+$ remains practically constant, such that its spectrum is dominated by a sharp maximum at zero frequency. In contrast, the time evolution of the out-of-phase mode is given by the phase factor $\Psi_- = e^{iB(t)}$, where $\dot B = \varepsilon(t)$. In Fourier space, it becomes the integral
\begin{equation}
\Psi_-(\omega) = \int dt \, e^{i\omega t+iB(t)}\,,
\label{Psi-FT}
\end{equation}
which we evaluate within the stationary phase approximation. This means that we replace the integral by its contributions in the vicinity of times at which the time derivative of the exponent vanishes, i.e., when the equation $\omega = -\dot B = -\varepsilon(t)$ is fulfilled.  As $\varepsilon$ is bounded, the spectrum is essentially restricted to the frequency range $-\max(\varepsilon) \leq \omega \leq -\min(\varepsilon)$. The contribution of each stationary point is proportional to $\sqrt{2\pi/\dot\varepsilon(t_n)}$, where $\dot\varepsilon$ is the second time derivative of the phase in Eq.~\eqref{Psi-FT}.  This time derivative vanishes at the extrema of $\varepsilon(t)$, such that the $\Psi_-(\omega)$ diverges at the margins of the spectrum. For the assumed shape of the driving, we expect pronounced maxima near $-\varepsilon_0\pm A$.

\begin{figure}
\centerline{\includegraphics{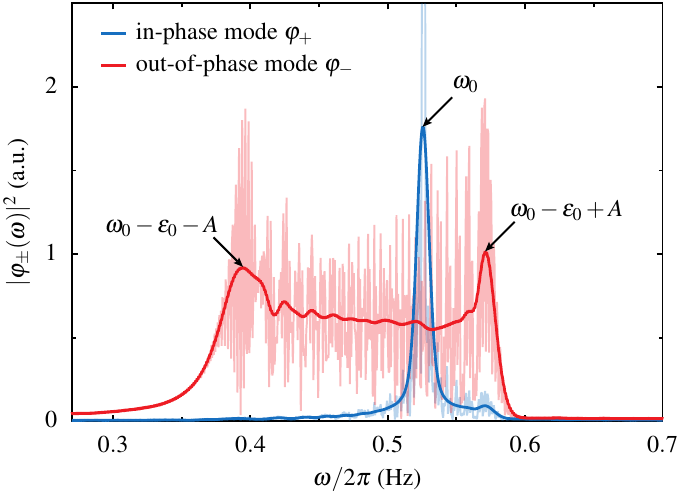}}
\caption{Spectrum of the in-phase and out-of-phase modes for a typical LZSM experiment. The faint lines are the numerical Fourier transforms of the raw data, while the thicker lines are the same data but smoothened with a Gaussian of width $\sigma=\SI{0.025}{\Hz}$. The maxima marked by arrows allow us to roughly estimate the detuning $\epsilon_0$ and the driving amplitude $A$ as discussed in the text.}
\label{fig:spectrum}
\end{figure}

There is no need to remove the phase factor $e^{-i\omega_0 t}$ as in the
ansatz \eqref{app:ansatz}, because it only shifts the Fourier spectrum to
higher frequencies.  Thus, in \fig{fig:spectrum}{} 
we performed the analysis directly with the spectra of $\varphi_\pm$ measured for a typical LZSM measurement. Thin lines correspond to the numerically Fourier transformed measured $\varphi_\pm$. While $\varphi_+$ yields a peak at the already known value $\omega_0$, the spectrum of $\varphi_-$ is broadened due to the driving and develops a frequency comb spectrum with two main maxima, one near $\omega_0-\varepsilon_0-A$ and a second one near $\omega_0-\varepsilon_0+A$. 
To determine these maxima we convolved the experimental spectra with a Gaussian resulting in the fat lines. The positions of the maxima of the out-of-phase mode provide a rough estimate of the parameters $\varepsilon_0$ and $A$.

Unfortunately, the positions of the maxima of the out-of-phase mode spectrum are not exactly at $\omega_0-\varepsilon_0\pm A$. In fact, this Fourier analysis method to determine $A$ and $\varepsilon_0$ systematically underestimates $A$. In principle, this systematic error could be corrected for in a calibration based on a comparison with alternative methods discussed below.

\subsubsection{Husimi analysis}

The next method is capable of determining the full time dependence of $\varepsilon(t)$ by employing a phase-space method frequently used for visualizing semi classical structures of a quantum mechanical wave function $\varphi(x)$.  It consists of a mapping of $\varphi(x)$ to a function $Q(x,p)$, whose structure marks the corresponding classical orbits $(x_t,p_t)$. As such it provides the momentum as a function of the position, $p(x)$, with a resolution limited by the uncertainty principle.

\begin{figure}
\centerline{\includegraphics{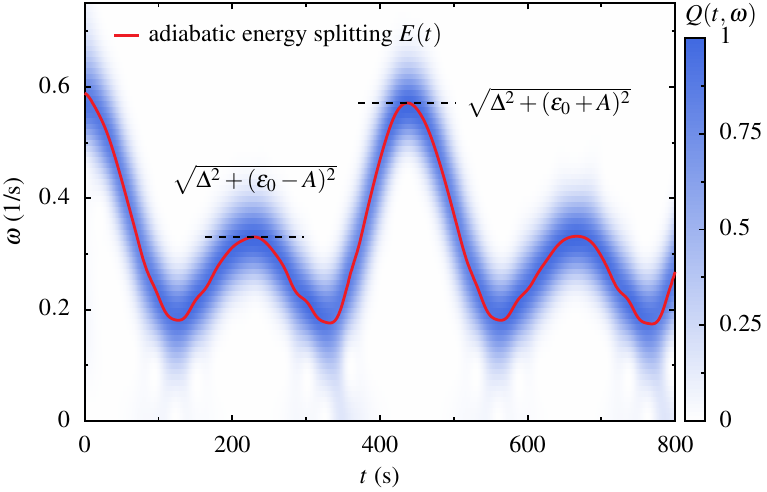}}
\caption{Husimi representation $Q(t)$ (blue scale) of the out-of-phase mode
$\Psi_-(t)$ with its spectrum shown in Fig.~\ref{fig:spectrum}.  The
average (red solid line) provides an estimate of the time evolution of the
splitting $E(t) = \sqrt{\Delta^2+\varepsilon(t)^2}$. It's maxima allow one
to estimate $\epsilon_0$ and $A$, see text.
}
\label{fig:husimi}
\end{figure}

Replacing the phase space coordinates $(x,p)$ by time and frequency, one obtains a mapping of a function of time, $\Psi(t)$, to $Q(t,\omega)$. Accordingly,  $Q(t,\omega)$ provides the time-resolved oscillation frequency $\omega(t)$.  For a two-level system described by the Hamiltonian \eqref{Htls}, the relevant frequency scale is the splitting between the eigenmodes $\omega(t)=E(t) = \sqrt{\Delta^2+\varepsilon(t)^2}$. Thus, while the time-frequency Husimi representation of $\Psi_+(t)$ is constant at zero, that of $\Psi_-(t)$ traces the adiabatic splitting $E(t)$. In \fig{fig:husimi}{} we plot the Husimi representation of the out-of-phase mode $\Psi_-(t)$ for the data shown in \fig{fig:spectrum}{}.

The Husimi function can be defined as the modulus squared of the overlap of a function with a wave packet centered at position $t$ and oscillating with a frequency $\omega$,
\begin{equation}
w_{t,\omega}(t') = \exp\Big({-\frac{(t-t')^2}{2\sigma^2} -i\omega t'}\Big) ,
\end{equation}
where the width $\sigma$ shifts the uncertainty towards time (large
$\sigma$) or frequency (small $\sigma$).
Thus, $Q(t,\omega) = |q(t,\omega)|^2$, where
\begin{equation}
\begin{split}
q(t,\omega) ={}& \int dt'\, w_{t,\omega}^*(t') \Psi(t')
\\
={}& e^{-i\omega t} \int dt'\, w_{0,\omega}^*(t-t') \Psi(t') .
\end{split}
\label{q}
\end{equation}
The convolution form obtained with the second line is convenient for the numerical evaluation, while the phase factor does not affect $Q(t,\omega)$.

The interpretation by which we motivated the use of the Husimi function becomes evident when one considers a function $\Psi(t) = e^{-iE(t)t}$ with some slowly varying function $E(t)$, such that $dE/dt$ can be neglected. Evaluating the integral in Eq.~\eqref{q} within steepest descent, we have to determine the stationary points at which the $t'$-derivative of the exponent vanishes.  Real and imaginary part of this condition read $t'=t$ and $\omega = E(t)$, which means that the structure of $Q(t,\omega)$ is indeed dominated by the momentary oscillation frequency.

The price of the Husimi analysis are the fundamental restrictions of its resolution resulting in a broadening of $E(t)$ in time. As the Fourier analysis above, also this method is based on the evaluation of a Fourier integral within a stationary-phase approximation. As a consequence, it equally suffers from an underestimation of the splitting $E(t)$ and at its turning points and, hence, from an underestimation of $A$.

The main benefit of the Husimi analysis is its ability to directly visualize the time evolution of the modulation of the splitting $E(t)$ and, hence, the time-dependent coupling strength $|\varepsilon(t)|=\sqrt{E(t)^2-\Delta^2}$, where the minimal splitting is given by the frequency difference $\Delta$.

\subsubsection{Analysis of the quasistatic solution}

\begin{figure}
\centerline{\includegraphics{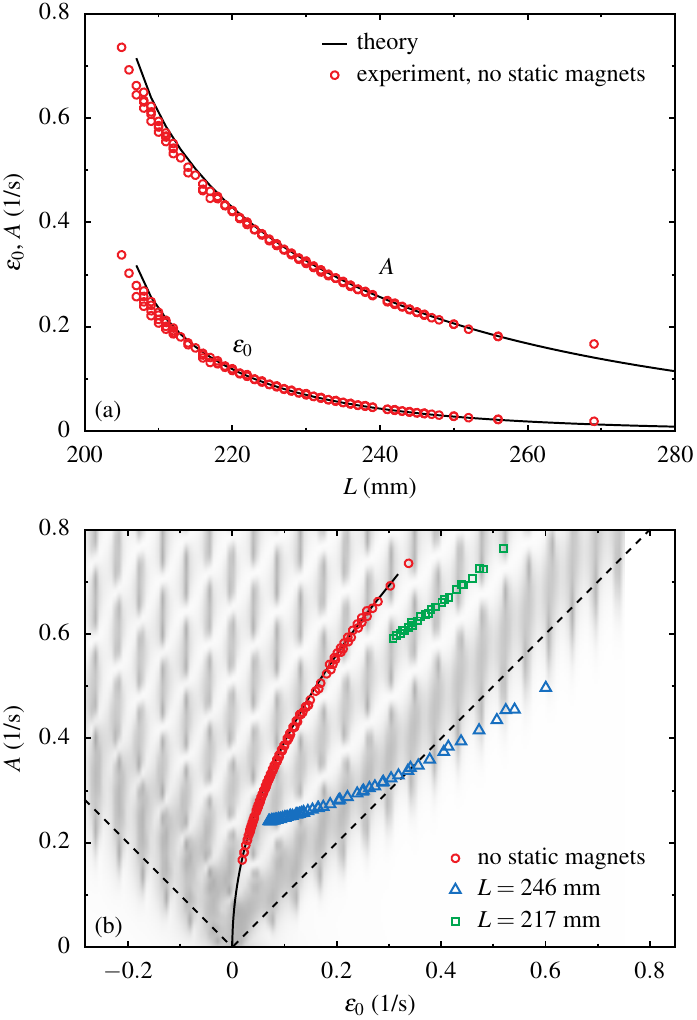}}
\caption{(a) Effective TLS modulation parameters, $A(L)$ and $\varepsilon_0(L)$, reconstructed from the LZSM dynamics without upper magnets for modulation period $T=\SI{441}{s}$. The red circles result from the analysis of the quasistatic solution $\varphi_{1,2}^\text{qs}(t)$ using \eq{A-eps}, while the solid lines corresponds to the numerical solution based on \eq{epsilonbar}.  The spectral analyses (Fourier or Husimi analyses, not shown) yield comparable results within their respective accuracies.
(b) LZSM interference pattern (gray scale) adopted from Fig.\ \ref{fig:lzsmfan}b calculated with the Schr\"odinger equation. The red circles depict $A(\varepsilon_0)$ without upper magnets determined from pairs of data points shown in panel a with varying $L$, the solid line shows the according prediction. Other symbols show $A(\varepsilon_0)$ for measurements with constant $L=217\,$mm (squares) or $L=246\,$mm (triangles) while we varied the distance $L_\text{u}$ between the now present upper magnets.
}
\label{fig:AEeff}
\end{figure}

To overcome such uncertainties, in our third method we determine $\varepsilon_0$ and $A$ directly from the measured quasistatic solution $\varphi_{1,2}^\text{qs}(t)$ discussed in \sect{sec:quasistatic}. It can be directly measured by rotating a magnet without exciting the pendula otherwise, such that $\varphi_{1,2}(t)=\varphi_{1,2}^\text{qs}(t)$. Alternatively, as $\Omega\ll\omega_0$, one can extract $\varphi_{1,2}^\text{qs}(t)$ with high accuracy from measurements with oscillating pendula by applying a digital lowpass to separate the quasistatic dynamics from $\varphi_{1,2}(t)$, cf.\ \sect{sec:processing}. 
Based on our approximation $\varepsilon=\varepsilon_0+A\cos(\Omega t)$ we determine the extreme values $\varepsilon_\text{min}$ and $\varepsilon_\text{max}$ and use
\begin{equation}
\begin{split}
A         &=\frac 12 \left(\varepsilon_\text{max}-\varepsilon_\text{min}\right)\\
\varepsilon_0&=\frac 12 \left(\varepsilon_\text{max}+\varepsilon_\text{min}\right)\;.
\end{split}
\label{A-eps}
\end{equation}
The symbols in \fig{fig:AEeff}{a} 
present the modulation parameters $\epsilon_0$ and $A$ of the effective TLS as a function of the pivot distance $L$ determined from a series of experiments without upper (static) magnets using \eq{A-eps}. The solid line corresponds to the numerical solution based on \eq{epsilonbar}. The agreement is excellent for $L\gtrsim220\,$mm, while for smaller distances we find noticeable deviations. They indicate, that for our stronger couplings a quantitative derivation of the interaction parameters of the effective TLS from the nonlinear dipole-dipole interaction has its limitations. For this reason, we use the experimental values $\epsilon_0(L)$ and $A(L)$ for our further analysis. 

In \fig{fig:AEeff}{b} we present as a grayscale the LZSM interference pattern already shown in \fig{fig:lzsmfan}b, which we calculated numerically based on the Schr\"odinger equation. On top we plot the values of $A(\varepsilon_0)$ determined from our experiments as symbols. The open circles correspond to the $\epsilon_0(L)$ and $A(L)$ values shown in \fig{fig:AEeff}a for the case of no upper magnets. The solid line behind these points is the prediction based on \eq{epsilonbar} and corresponds to the solid lines in \fig{fig:AEeff}{a}. Triangles and squares show experimental values for measurements including upper magnets. In these experiments the distance between the pivots was constant while the distance between the upper magnets was varied. These $A(\varepsilon_0)$ curves are the basis for the comparison of the measured and predicted LZSM interference patterns presented in \fig{fig:LZSM}{c} and \fig{fig:LZSM}{d}.

\begin{figure}[t]
\centerline{\includegraphics{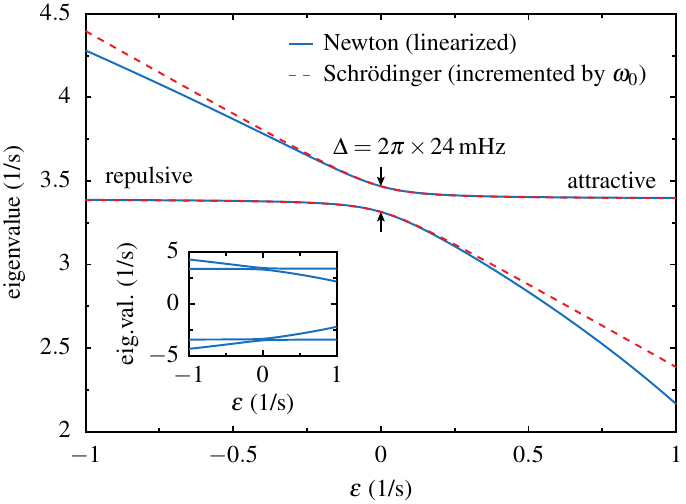}}
\caption{Avoided crossing formed by the eigenfrequencies of the coupled pendula calculated for the detuning of $f_1-f_2=24\,$mHz. The solid lines are computed with the linearized Newton's equations, the dashed lines are the eigenvalues of the Schr\"odinger equation. The inset displays the full spectrum of Newton's equations, including the negative eigenvalues.}
\label{fig:avoided_crossing}
\end{figure}

\subsubsection{Consistency of our approximations}

Our approximations are based on the following concept: Small oscillations of a classical many-body system can be described by a linearized equation of motion of the form $M\ddot{\bm{x}} = V\bm{x}$, where the coordinate vector $\bm{x}$ consists of all deviations from the equilibrium position.  The symmetric matrix $V$ contains the potential curvatures, while $M$ is the diagonal matrix of the masses of each particle.  Since the masses are positive, $M^{1/2}$ is real valued, such that the equation of motion can be written as $\ddot{\bm{y}} = -Q\bm{y}$ with the symmetric matrix $Q = M^{-1/2} V M^{-1/2}$, which has real and non-negative eigenvalues (see Ref.~\cite{Goldstein2001} or another text book on classical mechanics).  Their square roots are the eigenvalues of the linearized equations of motion and are shown in the inset of \fig{fig:avoided_crossing}{}.

As a consistency check for our mapping to a Schr\"odinger equation we verify that its spectrum corresponds to the one of the linearized classical equation of motion.  Since we neglect the lower half of the spectrum in the rotating-wave approximation, our results compare to the positive eigenfrequencies of Newton's equation. In addition, our ansatz \eqref{app:ansatz} corresponds to a gauge transformation that shifts the eigenfrequencies by $-\omega_0$.  Hence, the spectrum predicted by our Schr\"odinger equation shifted by $\omega_0$ finally corresponds to the positive eigenfrequencies of Newton's equation. In the main panel of \fig{fig:avoided_crossing}{} we present both spectra in direct comparison in the absence of the driving. For sufficiently small interaction $\varepsilon$ our mapping is accurate. We expect, that the mapping works equally well for our driven experiments, as we consider slow driving with $\Omega\ll\omega_0$.

\section{Experimental setup and methods}
\label{app:experiment}

For visualizing the coherent wave mechanics equivalent to the dynamics of an individual qubit, macroscopic pendula have decisive advantages. First, in contrast to nanoscale devices, our pendula are large and have a slow clock speed, such that their dynamics can be observed with bare eyes. Second, the time evolution of a classical and macroscopic device can be obtained from one single experiment, while quantum- or nanosystems require many repeated measurements at various times. A drawback of our macroscopic pendula is, that a continuous modulation of the eigenfrequencies, corresponding to the energy detuning typically modulated in a qubit, is virtually impossible. It would require moving a weight smoothly up and down, for some measurements along the full length of a pendulum rod while it is oscillating. Therefore, in our experiments we keep the frequency detuning fixed and instead modulate the coupling between the pendula. Modulating the detuning or the coupling are mathematically equivalent options, which can be demonstrated by performing a unitary basis transformation in \sect{sec:basistransformation}. The in-phase and out-of-phase modes of our coupled pendula then correspond to the diabatic (or localized) states of a qubit, while the individual pendula eigenfrequencies correspond to its adiabatic eigenstates, cf.\ Table~\ref{table:params}. 

In order to be able to modulate the coupling, we replace the usual spring connecting both pendula by permanent magnets connected to each pendulum. We then modulate the coupling by rotating one of the magnets with a constant angular frequency. The setup is presented in \fig{fig:model}{} and can be experienced in the attached movie. Our magnets are cubicles of pressed neodymium powder coated with nickel bought from Webcraft GmbH (www.supermagnete.de).

\subsection{Details of the setup}

Each pendulum consists of a one meter long stainless steel rod with a diameter of 12\,mm, extended at the bottom end with a stainless steel thread and a hollow polyethylen housing containing AA batteries, which can drive a linear rotating motor via a simple circuit board. Two cubic neodymium magnets with 28\,mm edge length are glued to the axes of each motor. In the experiments discussed here, we rotate one of the two magnets with constant angular frequency. At a distance of \SI{0.513}{m} above them are two smaller magnets fixed by plastic screws inside plastic cylinders [red in the photograph in \fig{fig:pendula2}b] to the rods. These magnets can be moved horizontally inside the cylinders. Brass nuts fixed to the opposite ends of the cylinders function as counter weights to balance the center of masses of the pendula within the respective rods. Heavy brass cylinders (2.14\,kg) are screwed onto the threads attached to the pendula. The vertical positions of these weights serve for adjusting the resonance frequencies of the individual pendula. The overall weight of each pendulum is 4.242\,kg. In our experiments the air friction can be neglected compared to the friction of the pivots and the damping related to magnetic induction. The frame supporting the pendula is built from hollow aluminum bars, while the pendula are fixed via their pivots to a massive pair of stainless steel beams spanning the top of the frame. The pivots are professional pendulum clock pivots based on leaf springs provided by the company Erwin Sattler GmbH \& Co.\ KG. Due to their plate geometry the leaf springs strongly suppress unwanted pendula motions others than oscillations in the $x$-$y$ plane. The pivots quality is essential for providing high enough and stable quality factors. It is important to run the experiment in a tranquil surrounding, because in particular air flows and vibrations can cause uncontrolled phase shifts of the pendula oscillations. Therefore we have placed the frame supporting the pendula on a massive granite plate in a separate and quiet room in the cellar of the building. The coupling between the pendula is provided by up to four magnets. It is modulated whenever one of the lower magnets is rotated. At the high quality factors of several thousands it is necessary to avoid even tiny contributions to the coupling between the pendula mediated by the supporting frame. Initially this was a problem in our setup, which we prevented by stiffening the frame by adding brackets in its corners and by tightly fixing the frame at one side to an approximately 0.5\,m thick brick wall of the building.

\subsection{Requirements, line width and strong coupling}

For performing qubit simulating experiments, such as Rabi oscillations or LZSM interferometry, the resonance frequencies of the individual pendula have to be much higher than both, the coupling constant and the frequency difference, while the latter two must be highly tunable. At the same time, the quality factor must be high enough to ensure that the coupling strength exceeds the line widths of the eigenmodes by far. Practically, it is desired that for the duration of the experiment damping effects can be neglected, which is the case in our experiments and greatly simplifies the data analysis. The strong distance dependence of the magnetic dipole-dipole interaction provides the desired tunability of the coupling via adjusting the mutual distances between the lower and the upper magnets. Further, rotating one of the magnets allows us to modulate the coupling of the ``qubit'' analogue. The price is a time dependent momentary equilibrium deflection of the pendulum rods, which is discussed in detail in Appendix \ref{app:newton}.

If two magnets are moved with respect to each other, their electric conductivity gives rise to eddy currents, which result in the main damping mechanism of our coupled pendula, similar to the functioning of an induction break. The damping is weak, as the $Q$-factor of our coupled pendula still ranges between $3000-6000$ depending on the average distance of the magnets. With oscillation frequencies $\omega_0/2\pi\sim0.5$\,Hz, it allows us to observe the qubit equivalent dynamics for several hours. More importantly, our large $Q$-factors allow us to ignore damping effects within a limited time window $\delta t\ll 2\pi Q/\omega_0$, which facilitates a one-to-one comparison with a quantum mechanical two-level system.   

To resolve the splitting of a two-level-system it needs to exceed the line widths $\gamma=\omega_0/2Q$ of the eigenmodes, $\sqrt{\Delta^2+\varepsilon^2}>\gamma$. For our $Q>3000$ we find $\gamma<0.6\,\text{ms}^{-1}$ in our experiments. The splitting between the eigenmodes can then be easilly resolved by using frequency detunings $|\Delta|>\gamma$. Moreover, for coupled pendula with such a high $Q$-factor it is straightforward to realize the so-called strong coupling regime defined by $\varepsilon>\gamma$, which is reached in most of our experiments.

For achieving a meaningful comparison between our classical system and a qubit we require a clean separation between the individual pendulum frequencies and all other time scales $\Omega,\Delta,A\ll\omega_0$. We used the modulation frequencies $\Omega/2\pi=2.3\,\text{mHz}$, $\Omega/2\pi=7.1\,\text{mHz}$ or $\Omega/2\pi=11.7\,\text{mHz}$, modulation amplitudes of the coupling between $0.7\,\text{mHz}\lesssim A/2\pi\lesssim43\,\text{mHz}$, and frequency detunings $|\Delta|/2\pi\le5\,\text{mHz}$. To adjust the latter, we re-positioned 2\,kg weights along the pendulum rods. 

\section{Measurement Regimes}
\label{app:regimes}

\subsection{Rabi experiments}

Rabi oscillations can be observed in the limit of small couplings and if the much larger frequency detuning is similar to the modulation frequency, $\Delta\sim\Omega\gg\omr=A/2$, cf.\ \sect{sec:Rabi}. For simplicity we performed our Rabi experiments without upper magnets, such that (for small couplings) $A\simeq\varepsilon_\text{max}$. Practically, our modulation frequencies of a few mHz dictate a range of useful frequency differences $\Delta$ and couplings $A<\Delta$, the latter being controlled by the distance $L$ between the pivots.

\begin{figure}[t]
\centerline{\includegraphics{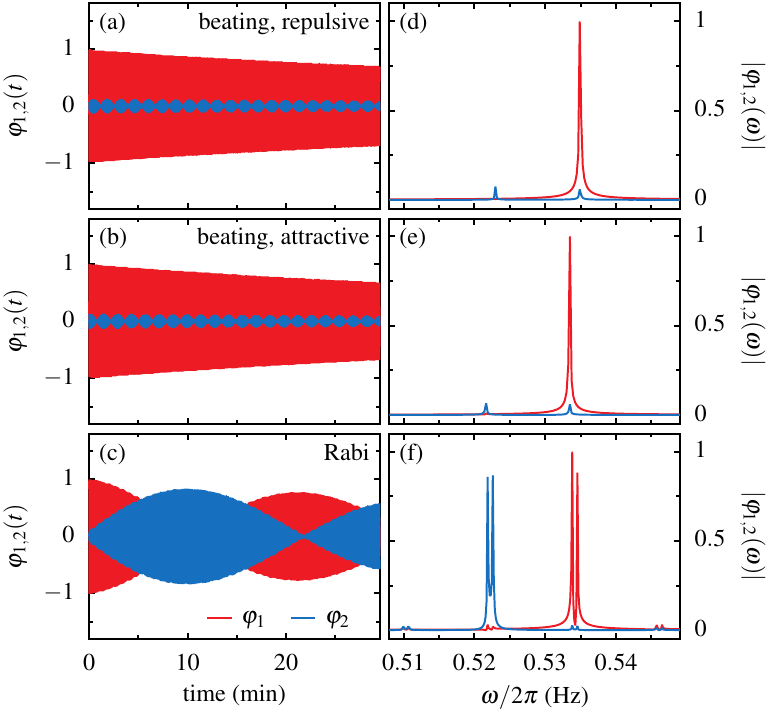}}
\caption{Time evolution of $\varphi_{1,2}$ for beating experiments with maximal repulsive (a) versus maximal attractive (b) couplings and the corresponding resonant Rabi experiment with the coupling modulated at the frequency $\Omega=\Delta$; $L=454\,$mm, $\Delta/2\pi=11.7\,$mHz. (d--f), Fourier transforms of the time evolutions shown in panels a--c.}
\label{fig:Rabi_comp}
\end{figure}
In \twofigs{fig:Rabi_comp}{a}{b} we present the deflections of both pendula $\varphi_k(t)$ for beating experiments without driving ($A=0$, $\varepsilon=\varepsilon_\text{min}$ or $\varepsilon=\varepsilon_\text{max}$) for the two extreme coupling cases with the magnets aligned either antiparallel for maximal repulsion or collinear for maximal attraction, where the pivot distance $L=454\,$mm corresponds to a small coupling. To initialize each measurement, we deflected just one of the two pendula, the one corresponding to the red lines. The energy transfer between the two pendula is clearly incomplete owing to the finite frequency detuning, $\Delta/2\pi=11.7\,$mHz, while the beating frequency is $\sqrt{\Delta^2+\varepsilon^2}/2\pi\simeq\Delta/2\pi$.  The latter corresponds to the difference between the respective eigenfrequencies, directly visible in the Fourier spectra shown in \twofigs{fig:Rabi_comp}{d}{e}. The spectrum of the pendulum that was initially not deflected (blue) clearly contains two maxima, where the frequency of the smaller peak coincides with the main maximum of the initially deflected pendulum (red). This indicates a finite mixing between the states represented by $\varphi_1$ and $\varphi_2$, which are not the exact eigenmodes because of the coupling between the pendula. Note, that the eigenfrequencies are slightly smaller for the attractive interaction as compared to the case of repulsive interaction.

In \fig{fig:Rabi_comp}{c} we present the corresponding resonant Rabi experiment with identical parameters as above but the coupling being modulated with the angular frequency $\Omega=\Delta$. In this case, the initial beating experiments mark the turning points of the modulation of the coupling during the Rabi experiment. The energy transfer between the two pendula is now complete but happens at the Rabi frequency $\omr/2\pi\simeq0.7\,\text{mHz}$, hence the initially postulated condition $\omr\ll\Delta,\Omega$ is fulfilled. 

The Fourier spectra of the Rabi experiment in \fig{fig:Rabi_comp}{f} reveal two main peaks and four side peaks for each pendulum. The main peaks are splitted by the (effective) Rabi frequency in \eq{eq:effRabi}. The much smaller side peaks, each of which is equally split by the (effective) Rabi frequency, are higher order components split off by multiples of $\Omega$ from the main peaks. In the resonant case $\Omega=\Delta$, the frequency values of the Fourier components of both pendula coincide, in the non-resonant case they would be displaced by $\Omega-\Delta$. Note that the higher order components in the Fourier spectra are responsible for the weak stepwise modulation with frequency $\Omega$ of the occupation of the pendula, which are weakly visible in \fig{fig:Rabi_comp}{c}.

\subsection{LZSM experiments}
\begin{figure}[t]
\centerline{\includegraphics{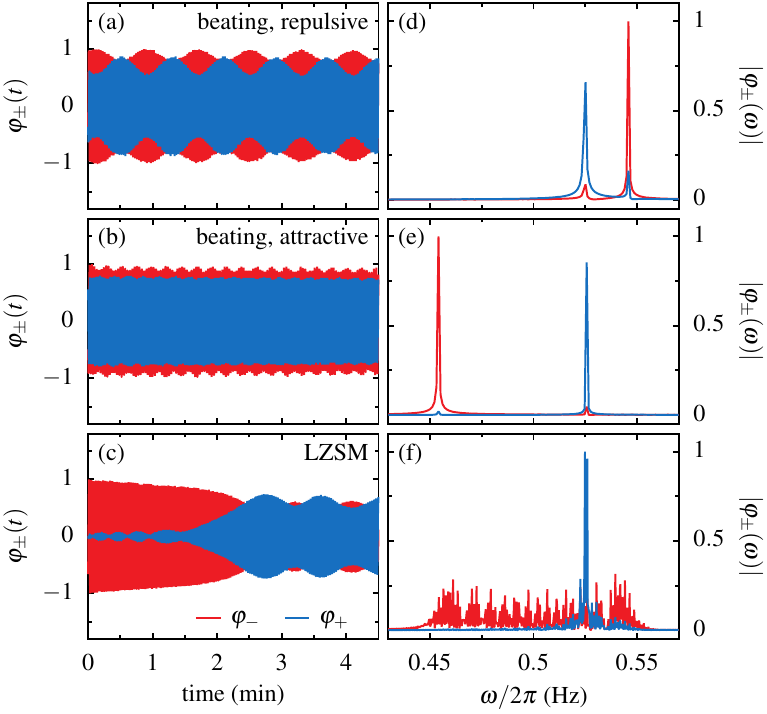}}
\caption{Time evolution of $\varphi_{\pm}$ for beating experiments with maximal repulsive (a) versus maximal attractive (b) couplings and the corresponding LZSM experiment with the coupling modulated at the frequency $\Omega/2\pi=2.27\,$mHz; $L=240\,$mm, $\Delta/2\pi=6.2\,$mHz. (d--f), Fourier transforms of the time evolutions shown in panels a--c; for the LZSM experiment the Fourier transform in panel f corresponds to five full periods of the modulation, while panel c shows only one period.}
\label{fig:LZSM_comp}
\end{figure}

For a direct comparison with the small coupling regime of Rabi experiments we perform similar measurements within the regime of LZSM experiments at much larger modulation of the coupling with $A>\Omega$. We consider measurements for $L=240\,$mm, $\Delta/2\pi=6.2\,$mHz, $\Omega/2\pi=2.27\,$mHz and a sizable $\varepsilon_0$ realized by including upper magnets.  For such large couplings ($A,|\varepsilon_0|\gg\Delta$) the diabatic modes $\varphi_\pm$ approximately correspond to the eigenmodes.  Hence, in \fig{fig:LZSM_comp}{} we now plot $\varphi_\pm$. The beating experiments, which we again performed for collinear versus antiparallel magnets, summarized in the upper four panels of \fig{fig:LZSM_comp}{} reveal the expected much larger range of couplings compared to the Rabi experiment. The very different beating frequencies for attractive versus repulsive interactions point to a sizable $\varepsilon_0$. Note, that the frequency (main component of Fourier spectrum) of $\varphi_+$ is almost identical for repulsive versus attractive coupling (blue in Figs.\ \ref{fig:LZSM_comp}{d} and \ref{fig:LZSM_comp}{e}), while the frequency of $\varphi_-$ (red) varies by roughly 20\,\%.

The LZSM experiment, presented by its first avoided crossing in \fig{fig:LZSM_comp}{c}, shows the expected energy transfer between the in-phase and out-of-phase modes near the avoided crossing. The additional faster beats vary in frequency related with the time dependence of $\varepsilon$. The Fourier spectrum plotted in \fig{fig:LZSM_comp}{f} comprises five modulation periods. It reveals that the in-phase mode stays at the frequency of the beating experiments, while the out-of-phase mode contains frequency components spanning a slightly larger region than that between the out-of-phase mode frequencies of the beating experiments. Note, that the apparent splitting, say $\delta\omega$, of the in-phase mode $\varphi_+$ in the Fourier spectrum of the LZSM experiment resembles a slight difference between the frequencies that the in-phase mode has between attractive versus repulsive beating experiments. As a result, the the out-of phase mode spectrum is composed of two copies of a frequency comb with the same relative shift $\delta\omega$, each one characterized by equally spaced peaks separate by the modulation frequency $\Omega$.

\begin{figure}
\centerline{\includegraphics{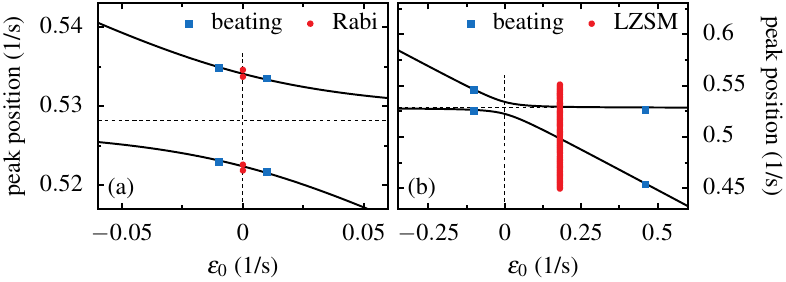}}
\caption{Avoided crossings (solid lines) with Fourier components of beating
experiments (blue squares, $\Omega=0$) and main Fourier components of modulated experiments (red circles, $\varepsilon=\varepsilon_0$).
(a) Rabi experiment for $f_1=0.53335\,$Hz, $f_2=0.52165\,$Hz, $L=454\,$mm and $\Omega/2\pi=11.7\,$mHz.
(b) LZSM experiment for $f_1=0.5290\,$Hz, $f_2=0.5228\,$Hz, $L=240\,$mm and $\Omega/2\pi=2.27\,$mHz.
}
\label{fig:Rabi_avoided_crossing}
\end{figure}

\subsection{Avoided crossings for Rabi versus LZSM experiments}

Figure \ref{fig:Rabi_avoided_crossing} summarizes the main components of the Fourier spectra of the experiments presented in Figs.~\ref{fig:Rabi_comp} and \ref{fig:LZSM_comp} and thereby visualizes the vastly different experimental regimes realized in a Rabi experiment, where $A<\Delta$, versus a LZSM experiment with $A\gg\Delta$. 

To highlight the differences between the regimes of the Rabi versus LZSM experiments, in \twofigs{fig:Rabi_avoided_crossing}{a}{b} we plot for the Rabi versus LZSM experiments shown in Figs.~\ref{fig:Rabi_comp} and \ref{fig:LZSM_comp} the relevant regions of the avoided crossings predicted for the measured frequencies by the Schr\"odinger equation (for the time independent quantum mechanical two-level system). The blue squares indicate the components of the respective Fourier spectra of the beating experiments, where we determined the values of $\varepsilon$ using the eigenvalue equation $\omega_\pm=\frac12(-\varepsilon\pm\sqrt{\Delta^2+\varepsilon^2})$. The red circles in \fig{fig:Rabi_avoided_crossing}{a} indicate the four main components of the Fourier spectra of the Rabi experiment in \fig{fig:Rabi_comp}{f}, where we used $\varepsilon=\varepsilon_0\simeq0$ for simplicity. The line of red circles in \fig{fig:Rabi_avoided_crossing}{b} indicates the range of the frequency comb of the Fourier spectrum of $\varphi_-$ of the LZSM experiment, cf.\ \fig{fig:LZSM_comp}{f}, where we again used $\varepsilon=\varepsilon_0$ for simplicity.

Clearly, for the presented Rabi experiment in \fig{fig:Rabi_avoided_crossing}{a}, $\varepsilon_0\simeq 0$ and $A\ll\Delta$, while for the LZSM experiment in \fig{fig:Rabi_avoided_crossing}{b}, both $\varepsilon_0,A > \Delta$. This depicts the main difference between the two regimes.

\section{Notations, units and magnitudes}
\label{app:notations}
\vspace{-1ex}
Tables \ref{table:params_individual}--\ref{table:params_driven} below summarize
the variables as well as their magnitudes used.

\begin{table*}[h!]
\caption{Variables of individual pendula}
\begin{ruledtabular}
\begin{tabular}{lll}\label{table:params_individual}
variable, definition & explanation & values
\\\hline\\[-1.5ex]
$\varphi_{1,2}$                     & deflection angles of individual pendula                & $|\varphi_{1,2}|<0.8^\circ\simeq0.014\,$rad
\\
$M$                                 & overall mass of each pendulum                          & 4.242\,kg
\\
$g$                                 & acceleration due to gravity in Munich (PTB table)      & 9.807232\,m/s$^2$
\\
$f_1$ $|$ $\omega_1=2\pi f_1$       & resonance $|$ angular frequency of pendulum 1          & $f_1\simeq(0.52 - 0.55)\,\text{Hz}$
\\
$f_2$ $|$ $\omega_2=2\pi f_2$       & resonance $|$ angular frequency of pendulum 2          & $f_2=0.52195\,\text{Hz}$
\\
$l_{\text r1}={g}/{\omega_1^2}$ & reduced length of pendulum 1                           & (0.818 - 0.912)\,m
\\
$l_{\text r2}={g}/{\omega_2^2}$ & reduced length of pendulum 2                           & 0.912\,m
\\
$l_\text{c1}$                       & center of mass of pendulum 1 (distance from pivot)     & $(0.754 - 0.841)\,$m
\\
$l_\text{c2}$                       & center of mass of pendulum 2 (distance from pivot)     & 0.841\,m
\\
$J_1=Ml_{\text c1}l_{\text r1}$     & moment of inertia of pendulum 1 (respective pivot)     & (2.619 - 3.254)\,kg\,m$^2$
\\
$J_2=Ml_{\text c2}l_{\text r2}$     & moment of inertia of pendulum 2 (respective pivot)     & 3.254\,kg\,m$^2$
\\
$Q_{1,2}$                           & quality factor of individual uncoupled pendula        & $9500-10500$
\\[1.0ex]
\multicolumn{2}{l}{The following distances are equal for both pendula:}                    &
\\ 
$l_\text{p}$                        & distance between pivot and point of measurement                       & 1.053\,m
\\
$l_\text{l}$                        & distance between pivot and center of lower magnet                     & 1.148\,m
\\
$l_\text{u}$                        & vertical distance between pivot and center of upper magnet            & 0.635\,m
\end{tabular}
\end{ruledtabular}
%
%
\caption{Variables related with the coupling between the pendula}
\begin{ruledtabular}
\begin{tabular}{lll}\label{table:params_coupled}
variable, definition & explanation & values and units
\\\hline\\[-1.5ex]
$\varphi_\pm=(\varphi_1\pm\varphi_2)/2$ & in-phase | out-of-phase mode                                                          & $<0.8^\circ$
\\
$L$                                     & distance between pivots, i.e., lower magnets for $\varphi_1=\varphi_2=0$              & $(0.205-0.454)\,$m
\\
$m_\text{l}$                            & magnetic moment of each lower magnet                                                  & 25.37\,Am$^2$
\\
$L_\text{u}$                            & distance between upper  magnets for $\varphi_1=\varphi_2=0$                           & $(0.105-0.168)\,$m
\\
$m_\text{u}$                            & magnetic moment of each upper magnet                                                  & 6.544\,Am$^2$
\\
$J_0=(J_1+J_2)/2$                       & mean moment of inertia                                                                & $\simeq 3$\,kg\,m$^2$
\\
$\omega_0=(\omega_1+\omega_2)/2$        & mean angular frequency of both pendula                                                & $(3.28 - 3.37)$\,s$^{-1}$
\\
$\Delta=\omega_1-\omega_2$              & angular frequency difference                                                          & $(0 - 29)\,\times10^{-3}\text{s}^{-1}$
\\
$\varepsilon = {G}/{\omega_0 J_0}$  & coupling constant                                                                     & ($-0.4 - 1.3$)\,s$^{-1}$
\\
$G(G_\text{l},G_\text{u})$              & effective potential curvature (see \sect{sec:3rd-order})                              &
\\
$G_\text{l}={6\mu_0m_\text{l}^2 l_\text{l}^2}/{\pi L^5}$           & interaction energy between lower magnets               & ($0.1 - 5.6$)\,J
\\
$G_\text{u}={6\mu_0m_\text{u}^2 l_\text{u}^2}/{\pi L_\text{u}^5}$  & interaction energy between upper magnets               & ($0.31-3.2$)\,J
\\
$Q$                                     & quality factor of coupled pendula (for realized range of $L$)           	            &  $3000-6000$
\\
\end{tabular}
\end{ruledtabular}
%
%
\caption{Additional variables related with the modulation of the coupling}
\begin{ruledtabular}
\begin{tabular}{lll}\label{table:params_driven}
variable, definition & explanation & values and units
\\\hline\\[-1.5ex]
$2\pi/\Omega$                                   & period of magnets' rotation                           & 85.5\,s, 141\,s, 441\,s
\\
$\varepsilon(t)=\varepsilon_0 + A\cos \Omega t$ & modulated coupling constant (linear approximation)    &  $\left[(-0.4)\, - (+1.3)\right]$\,s$^{-1}$
\\
$\omr=G_\text{l}/{2\omega_0 J_0}=A/2$ & rabi frequency for $\omr\ll\omega_0,\Delta$ and if only lower magnets are used & $(4.5-265)\,\times10^{-3}\text{s}^{-1}$
\\
$P_{\rm LZ}=\exp(-{\pi \Delta^2}/{2 |v|})$ & Single passage Landau-Zener probability     & $0 - 1$
\\
$v=\frac{d \varepsilon(t)}{d t}\big|_{\varepsilon=0}=\pm\Omega\sqrt{A^2-\varepsilon_0^2}$  & Speed of driving at avoided crossing at $\varepsilon=0$ & ($\pm0.0027 - \pm0.0300$)\,s$^{-2}$
\\
$P_{0}$                                         & Initial probability to occupy in-phase mode           & $0 - 0.2$
\\
\end{tabular}
\end{ruledtabular}
\end{table*}

\clearpage


\begin{thebibliography}{48}%
\makeatletter
\providecommand \@ifxundefined [1]{%
 \@ifx{#1\undefined}
}%
\providecommand \@ifnum [1]{%
 \ifnum #1\expandafter \@firstoftwo
 \else \expandafter \@secondoftwo
 \fi
}%
\providecommand \@ifx [1]{%
 \ifx #1\expandafter \@firstoftwo
 \else \expandafter \@secondoftwo
 \fi
}%
\providecommand \natexlab [1]{#1}%
\providecommand \enquote  [1]{``#1''}%
\providecommand \bibnamefont  [1]{#1}%
\providecommand \bibfnamefont [1]{#1}%
\providecommand \citenamefont [1]{#1}%
\providecommand \href@noop [0]{\@secondoftwo}%
\providecommand \href [0]{\begingroup \@sanitize@url \@href}%
\providecommand \@href[1]{\@@startlink{#1}\@@href}%
\providecommand \@@href[1]{\endgroup#1\@@endlink}%
\providecommand \@sanitize@url [0]{\catcode `\\12\catcode `\$12\catcode
  `\&12\catcode `\#12\catcode `\^12\catcode `\_12\catcode `\%12\relax}%
\providecommand \@@startlink[1]{}%
\providecommand \@@endlink[0]{}%
\providecommand \url  [0]{\begingroup\@sanitize@url \@url }%
\providecommand \@url [1]{\endgroup\@href {#1}{\urlprefix }}%
\providecommand \urlprefix  [0]{URL }%
\providecommand \Eprint [0]{\href }%
\providecommand \doibase [0]{https://doi.org/}%
\providecommand \selectlanguage [0]{\@gobble}%
\providecommand \bibinfo  [0]{\@secondoftwo}%
\providecommand \bibfield  [0]{\@secondoftwo}%
\providecommand \translation [1]{[#1]}%
\providecommand \BibitemOpen [0]{}%
\providecommand \bibitemStop [0]{}%
\providecommand \bibitemNoStop [0]{.\EOS\space}%
\providecommand \EOS [0]{\spacefactor3000\relax}%
\providecommand \BibitemShut  [1]{\csname bibitem#1\endcsname}%
\let\auto@bib@innerbib\@empty
\bibitem [{\citenamefont {Shore}\ \emph {et~al.}(2009)\citenamefont {Shore},
  \citenamefont {Gromovyy}, \citenamefont {Yatsenko},\ and\ \citenamefont
  {Romanenko}}]{ShoreAJP09}%
  \BibitemOpen
  \bibfield  {author} {\bibinfo {author} {\bibfnamefont {B.~W.}\ \bibnamefont
  {Shore}}, \bibinfo {author} {\bibfnamefont {M.~V.}\ \bibnamefont {Gromovyy}},
  \bibinfo {author} {\bibfnamefont {L.~P.}\ \bibnamefont {Yatsenko}},\ and\
  \bibinfo {author} {\bibfnamefont {V.~I.}\ \bibnamefont {Romanenko}},\
  }\bibfield  {title} {\bibinfo {title} {Simple mechanical analogs of rapid
  adiabatic passage in atomic physics},\ }\href
  {https://doi.org/10.1119/1.3231688} {\bibfield  {journal} {\bibinfo
  {journal} {Am. J. Phys.}\ }\textbf {\bibinfo {volume} {77}},\ \bibinfo
  {pages} {1183} (\bibinfo {year} {2009})}\BibitemShut {NoStop}%
\bibitem [{\citenamefont {Rabi}(1937)}]{RabiPR37}%
  \BibitemOpen
  \bibfield  {author} {\bibinfo {author} {\bibfnamefont {I.~I.}\ \bibnamefont
  {Rabi}},\ }\bibfield  {title} {\bibinfo {title} {Space quantization in a
  gyrating magnetic field},\ }\href {https://doi.org/10.1103/PhysRev.51.652}
  {\bibfield  {journal} {\bibinfo  {journal} {Phys. Rev.}\ }\textbf {\bibinfo
  {volume} {51}},\ \bibinfo {pages} {652} (\bibinfo {year} {1937})}\BibitemShut
  {NoStop}%
\bibitem [{\citenamefont {Landau}(1932)}]{LandauPZS32a}%
  \BibitemOpen
  \bibfield  {author} {\bibinfo {author} {\bibfnamefont {L.~D.}\ \bibnamefont
  {Landau}},\ }\bibfield  {title} {\bibinfo {title} {Zur {T}heorie der
  {E}nergie\"ubertragung bei {S}t\"o{\ss}en},\ }\href@noop {} {\bibfield
  {journal} {\bibinfo  {journal} {Phys. Z. Sowjetunion}\ }\textbf {\bibinfo
  {volume} {2}},\ \bibinfo {pages} {46} (\bibinfo {year} {1932})}\BibitemShut
  {NoStop}%
\bibitem [{\citenamefont {Zener}(1932)}]{ZenerPRSLA32}%
  \BibitemOpen
  \bibfield  {author} {\bibinfo {author} {\bibfnamefont {C.}~\bibnamefont
  {Zener}},\ }\bibfield  {title} {\bibinfo {title} {Non-adiabatic crossing of
  energy levels},\ }\href {https://doi.org/10.1098/rspa.1932.0165} {\bibfield
  {journal} {\bibinfo  {journal} {Proc. R. Soc. London A}\ }\textbf {\bibinfo
  {volume} {137}},\ \bibinfo {pages} {696} (\bibinfo {year}
  {1932})}\BibitemShut {NoStop}%
\bibitem [{\citenamefont {Stueckelberg}(1932)}]{StueckelbergHPA32}%
  \BibitemOpen
  \bibfield  {author} {\bibinfo {author} {\bibfnamefont {E.~C.~G.}\
  \bibnamefont {Stueckelberg}},\ }\bibfield  {title} {\bibinfo {title} {Theorie
  der unelastischen {S}t\"osse zwischen {A}tomen},\ }\href@noop {} {\bibfield
  {journal} {\bibinfo  {journal} {Helv. Phys. Acta}\ }\textbf {\bibinfo
  {volume} {5}},\ \bibinfo {pages} {369} (\bibinfo {year} {1932})}\BibitemShut
  {NoStop}%
\bibitem [{\citenamefont {Majorana}(1932)}]{MajoranaNC32}%
  \BibitemOpen
  \bibfield  {author} {\bibinfo {author} {\bibfnamefont {E.}~\bibnamefont
  {Majorana}},\ }\bibfield  {title} {\bibinfo {title} {Atomi orientati in campo
  magnetico variable},\ }\href@noop {} {\bibfield  {journal} {\bibinfo
  {journal} {Nuovo Cimento}\ }\textbf {\bibinfo {volume} {9}},\ \bibinfo
  {pages} {43} (\bibinfo {year} {1932})}\BibitemShut {NoStop}%
\bibitem [{\citenamefont {Gr{\o}nbech-Jensen}\ and\ \citenamefont
  {Cirillo}(2005)}]{GronbechJensenPRL05}%
  \BibitemOpen
  \bibfield  {author} {\bibinfo {author} {\bibfnamefont {N.}~\bibnamefont
  {Gr{\o}nbech-Jensen}}\ and\ \bibinfo {author} {\bibfnamefont
  {M.}~\bibnamefont {Cirillo}},\ }\bibfield  {title} {\bibinfo {title}
  {Rabi-type oscillations in a classical {J}osephson junction},\ }\href
  {https://doi.org/10.1103/PhysRevLett.95.067001} {\bibfield  {journal}
  {\bibinfo  {journal} {Phys. Rev. Lett.}\ }\textbf {\bibinfo {volume} {95}},\
  \bibinfo {pages} {067001} (\bibinfo {year} {2005})}\BibitemShut {NoStop}%
\bibitem [{\citenamefont {Novotny}(2010)}]{NovotnyAJP10}%
  \BibitemOpen
  \bibfield  {author} {\bibinfo {author} {\bibfnamefont {L.}~\bibnamefont
  {Novotny}},\ }\bibfield  {title} {\bibinfo {title} {Strong coupling, energy
  splitting, and level crossings: A classical perspective},\ }\href
  {https://doi.org/10.1119/1.3471177} {\bibfield  {journal} {\bibinfo
  {journal} {Am. J. Phys.}\ }\textbf {\bibinfo {volume} {78}},\ \bibinfo
  {pages} {1199–1202} (\bibinfo {year} {2010})}\BibitemShut {NoStop}%
\bibitem [{\citenamefont {Heinrich}\ \emph {et~al.}(2010)\citenamefont
  {Heinrich}, \citenamefont {Harris},\ and\ \citenamefont
  {Marquardt}}]{HeinrichPRA10}%
  \BibitemOpen
  \bibfield  {author} {\bibinfo {author} {\bibfnamefont {G.}~\bibnamefont
  {Heinrich}}, \bibinfo {author} {\bibfnamefont {J.~G.~E.}\ \bibnamefont
  {Harris}},\ and\ \bibinfo {author} {\bibfnamefont {F.}~\bibnamefont
  {Marquardt}},\ }\bibfield  {title} {\bibinfo {title} {Photon shuttle:
  {L}andau-{Z}ener-{S}t\"uckelberg dynamics in an optomechanical system},\
  }\href {https://doi.org/10.1103/PhysRevA.81.011801} {\bibfield  {journal}
  {\bibinfo  {journal} {Phys. Rev. A}\ }\textbf {\bibinfo {volume} {81}},\
  \bibinfo {pages} {011801(R)} (\bibinfo {year} {2010})}\BibitemShut {NoStop}%
\bibitem [{\citenamefont {Frimmer}\ and\ \citenamefont
  {Novotny}(2014)}]{FrimmerAJP14}%
  \BibitemOpen
  \bibfield  {author} {\bibinfo {author} {\bibfnamefont {M.}~\bibnamefont
  {Frimmer}}\ and\ \bibinfo {author} {\bibfnamefont {L.}~\bibnamefont
  {Novotny}},\ }\bibfield  {title} {\bibinfo {title} {The classical {B}loch
  equations},\ }\href {https://doi.org/10.1119/1.4878621} {\bibfield  {journal}
  {\bibinfo  {journal} {Am. J. Phys.}\ }\textbf {\bibinfo {volume} {82}},\
  \bibinfo {pages} {947–954} (\bibinfo {year} {2014})}\BibitemShut {NoStop}%
\bibitem [{\citenamefont {Ivakhnenko}\ \emph {et~al.}(2018)\citenamefont
  {Ivakhnenko}, \citenamefont {Shevchenko},\ and\ \citenamefont
  {Nori}}]{IvakhnenkoSR18}%
  \BibitemOpen
  \bibfield  {author} {\bibinfo {author} {\bibfnamefont {O.~V.}\ \bibnamefont
  {Ivakhnenko}}, \bibinfo {author} {\bibfnamefont {S.~N.}\ \bibnamefont
  {Shevchenko}},\ and\ \bibinfo {author} {\bibfnamefont {F.}~\bibnamefont
  {Nori}},\ }\bibfield  {title} {\bibinfo {title} {Simulating quantum dynamical
  phenomena using classical oscillators:
  {L}andau-{Z}ener-{S}t\"uckelberg-{M}ajorana interferometry, latching
  modulation, and motional averaging},\ }\href
  {https://doi.org/10.1038/s41598-018-28993-8} {\bibfield  {journal} {\bibinfo
  {journal} {Sci. Rep.}\ ,\ \bibinfo {pages} {12218}} (\bibinfo {year}
  {2018})}\BibitemShut {NoStop}%
\bibitem [{\citenamefont {Parafilo}\ and\ \citenamefont
  {Kiselev}(2018)}]{ParafiloPRB18}%
  \BibitemOpen
  \bibfield  {author} {\bibinfo {author} {\bibfnamefont {A.~V.}\ \bibnamefont
  {Parafilo}}\ and\ \bibinfo {author} {\bibfnamefont {M.~N.}\ \bibnamefont
  {Kiselev}},\ }\bibfield  {title} {\bibinfo {title} {Tunable {RKKY}
  interaction in a double quantum dot nanoelectromechanical device},\ }\href
  {https://doi.org/10.1103/PhysRevB.97.035418} {\bibfield  {journal} {\bibinfo
  {journal} {Phys. Rev. B}\ }\textbf {\bibinfo {volume} {97}},\ \bibinfo
  {pages} {035418} (\bibinfo {year} {2018})}\BibitemShut {NoStop}%
\bibitem [{\citenamefont {S\"usstrunk}\ and\ \citenamefont
  {Huber}(2015)}]{SusstrunkS15}%
  \BibitemOpen
  \bibfield  {author} {\bibinfo {author} {\bibfnamefont {R.}~\bibnamefont
  {S\"usstrunk}}\ and\ \bibinfo {author} {\bibfnamefont {S.~D.}\ \bibnamefont
  {Huber}},\ }\bibfield  {title} {\bibinfo {title} {Observation of phononic
  helical edge states in a mechanical topological insulator},\ }\href
  {https://doi.org/10.1126/science.aab0239} {\bibfield  {journal} {\bibinfo
  {journal} {Science}\ }\textbf {\bibinfo {volume} {349}},\ \bibinfo {pages}
  {47–50} (\bibinfo {year} {2015})}\BibitemShut {NoStop}%
\bibitem [{\citenamefont {Nash}\ \emph {et~al.}(2015)\citenamefont {Nash},
  \citenamefont {Kleckner}, \citenamefont {Read}, \citenamefont {Vitelli},
  \citenamefont {Turner},\ and\ \citenamefont {Irvine}}]{NashPNAS15}%
  \BibitemOpen
  \bibfield  {author} {\bibinfo {author} {\bibfnamefont {L.~M.}\ \bibnamefont
  {Nash}}, \bibinfo {author} {\bibfnamefont {D.}~\bibnamefont {Kleckner}},
  \bibinfo {author} {\bibfnamefont {A.}~\bibnamefont {Read}}, \bibinfo {author}
  {\bibfnamefont {V.}~\bibnamefont {Vitelli}}, \bibinfo {author} {\bibfnamefont
  {A.~M.}\ \bibnamefont {Turner}},\ and\ \bibinfo {author} {\bibfnamefont
  {W.~T.~M.}\ \bibnamefont {Irvine}},\ }\bibfield  {title} {\bibinfo {title}
  {Topological mechanics of gyroscopic metamaterials},\ }\href
  {https://doi.org/10.1073/pnas.1507413112} {\bibfield  {journal} {\bibinfo
  {journal} {Proc. Nat. Acad. Sci.}\ }\textbf {\bibinfo {volume} {112}},\
  \bibinfo {pages} {14495–14500} (\bibinfo {year} {2015})}\BibitemShut
  {NoStop}%
\bibitem [{\citenamefont {Faust}\ \emph {et~al.}(2013)\citenamefont {Faust},
  \citenamefont {Rieger}, \citenamefont {Seitner}, \citenamefont {Kotthaus},\
  and\ \citenamefont {Weig}}]{FaustNP13}%
  \BibitemOpen
  \bibfield  {author} {\bibinfo {author} {\bibfnamefont {T.}~\bibnamefont
  {Faust}}, \bibinfo {author} {\bibfnamefont {J.}~\bibnamefont {Rieger}},
  \bibinfo {author} {\bibfnamefont {M.~J.}\ \bibnamefont {Seitner}}, \bibinfo
  {author} {\bibfnamefont {J.~P.}\ \bibnamefont {Kotthaus}},\ and\ \bibinfo
  {author} {\bibfnamefont {E.~M.}\ \bibnamefont {Weig}},\ }\bibfield  {title}
  {\bibinfo {title} {Coherent control of a classical nanomechanical two-level
  system},\ }\href {https://doi.org/10.1038/nphys2666} {\bibfield  {journal}
  {\bibinfo  {journal} {Nature Phys.}\ }\textbf {\bibinfo {volume} {9}},\
  \bibinfo {pages} {485} (\bibinfo {year} {2013})}\BibitemShut {NoStop}%
\bibitem [{\citenamefont {Seitner}\ \emph {et~al.}(2016)\citenamefont
  {Seitner}, \citenamefont {Ribeiro}, \citenamefont {K\"olbl}, \citenamefont
  {Faust}, \citenamefont {Kotthaus},\ and\ \citenamefont
  {Weig}}]{SeitnerPRB16}%
  \BibitemOpen
  \bibfield  {author} {\bibinfo {author} {\bibfnamefont {M.~J.}\ \bibnamefont
  {Seitner}}, \bibinfo {author} {\bibfnamefont {H.}~\bibnamefont {Ribeiro}},
  \bibinfo {author} {\bibfnamefont {J.}~\bibnamefont {K\"olbl}}, \bibinfo
  {author} {\bibfnamefont {T.}~\bibnamefont {Faust}}, \bibinfo {author}
  {\bibfnamefont {J.~P.}\ \bibnamefont {Kotthaus}},\ and\ \bibinfo {author}
  {\bibfnamefont {E.~M.}\ \bibnamefont {Weig}},\ }\bibfield  {title} {\bibinfo
  {title} {Classical {S}t\"uckelberg interferometry of a nanomechanical
  two-mode system at room temperature},\ }\href
  {https://doi.org/10.1103/PhysRevB.94.245406} {\bibfield  {journal} {\bibinfo
  {journal} {Phys. Rev. B}\ }\textbf {\bibinfo {volume} {94}},\ \bibinfo
  {pages} {245406} (\bibinfo {year} {2016})}\BibitemShut {NoStop}%
\bibitem [{\citenamefont {Mullen}\ \emph {et~al.}(1989)\citenamefont {Mullen},
  \citenamefont {Ben-Jacob}, \citenamefont {Gefen},\ and\ \citenamefont
  {Schuss}}]{MullenPRL89}%
  \BibitemOpen
  \bibfield  {author} {\bibinfo {author} {\bibfnamefont {K.}~\bibnamefont
  {Mullen}}, \bibinfo {author} {\bibfnamefont {E.}~\bibnamefont {Ben-Jacob}},
  \bibinfo {author} {\bibfnamefont {Y.}~\bibnamefont {Gefen}},\ and\ \bibinfo
  {author} {\bibfnamefont {Z.}~\bibnamefont {Schuss}},\ }\bibfield  {title}
  {\bibinfo {title} {Time of {Z}ener tunneling},\ }\href
  {https://doi.org/10.1103/PhysRevLett.62.2543} {\bibfield  {journal} {\bibinfo
   {journal} {Phys. Rev. Lett.}\ }\textbf {\bibinfo {volume} {62}},\ \bibinfo
  {pages} {2543} (\bibinfo {year} {1989})}\BibitemShut {NoStop}%
\bibitem [{\citenamefont {Vitanov}(1999)}]{VitanovPRA99}%
  \BibitemOpen
  \bibfield  {author} {\bibinfo {author} {\bibfnamefont {N.~V.}\ \bibnamefont
  {Vitanov}},\ }\bibfield  {title} {\bibinfo {title} {Transition times in the
  {L}andau-{Z}ener model},\ }\href {https://doi.org/10.1103/PhysRevA.59.988}
  {\bibfield  {journal} {\bibinfo  {journal} {Phys. Rev. A}\ }\textbf {\bibinfo
  {volume} {59}},\ \bibinfo {pages} {988} (\bibinfo {year} {1999})}\BibitemShut
  {NoStop}%
\bibitem [{\citenamefont {Wubs}\ \emph {et~al.}(2005)\citenamefont {Wubs},
  \citenamefont {Saito}, \citenamefont {Kohler}, \citenamefont {Kayanuma},\
  and\ \citenamefont {H\"anggi}}]{WubsNJP05}%
  \BibitemOpen
  \bibfield  {author} {\bibinfo {author} {\bibfnamefont {M.}~\bibnamefont
  {Wubs}}, \bibinfo {author} {\bibfnamefont {K.}~\bibnamefont {Saito}},
  \bibinfo {author} {\bibfnamefont {S.}~\bibnamefont {Kohler}}, \bibinfo
  {author} {\bibfnamefont {Y.}~\bibnamefont {Kayanuma}},\ and\ \bibinfo
  {author} {\bibfnamefont {P.}~\bibnamefont {H\"anggi}},\ }\bibfield  {title}
  {\bibinfo {title} {{L}andau-{Z}ener transitions in qubits controlled by
  electromagnetic fields},\ }\href
  {https://doi.org/10.1088/1367-2630/10/21/218} {\bibfield  {journal} {\bibinfo
   {journal} {New J. Phys.}\ }\textbf {\bibinfo {volume} {7}},\ \bibinfo
  {pages} {218} (\bibinfo {year} {2005})}\BibitemShut {NoStop}%
\bibitem [{\citenamefont {Sillanp\"a\"a}\ \emph {et~al.}(2005)\citenamefont
  {Sillanp\"a\"a}, \citenamefont {Lehtinen}, \citenamefont {Paila},
  \citenamefont {Makhlin}, \citenamefont {Roschier},\ and\ \citenamefont
  {Hakonen}}]{SillanpaaPRL05}%
  \BibitemOpen
  \bibfield  {author} {\bibinfo {author} {\bibfnamefont {M.~A.}\ \bibnamefont
  {Sillanp\"a\"a}}, \bibinfo {author} {\bibfnamefont {T.}~\bibnamefont
  {Lehtinen}}, \bibinfo {author} {\bibfnamefont {A.}~\bibnamefont {Paila}},
  \bibinfo {author} {\bibfnamefont {Y.}~\bibnamefont {Makhlin}}, \bibinfo
  {author} {\bibfnamefont {L.}~\bibnamefont {Roschier}},\ and\ \bibinfo
  {author} {\bibfnamefont {P.~J.}\ \bibnamefont {Hakonen}},\ }\bibfield
  {title} {\bibinfo {title} {Direct observation of {J}osephson capacitance},\
  }\href {https://doi.org/10.1103/PhysRevLett.95.206806} {\bibfield  {journal}
  {\bibinfo  {journal} {Phys. Rev. Lett.}\ }\textbf {\bibinfo {volume} {95}},\
  \bibinfo {pages} {206806} (\bibinfo {year} {2005})}\BibitemShut {NoStop}%
\bibitem [{\citenamefont {Berns}\ \emph {et~al.}(2008)\citenamefont {Berns},
  \citenamefont {Rudner}, \citenamefont {Valenzuela}, \citenamefont {Berggren},
  \citenamefont {Oliver}, \citenamefont {Levitov},\ and\ \citenamefont
  {Orlando}}]{BernsNL08}%
  \BibitemOpen
  \bibfield  {author} {\bibinfo {author} {\bibfnamefont {D.~M.}\ \bibnamefont
  {Berns}}, \bibinfo {author} {\bibfnamefont {M.~S.}\ \bibnamefont {Rudner}},
  \bibinfo {author} {\bibfnamefont {S.~O.}\ \bibnamefont {Valenzuela}},
  \bibinfo {author} {\bibfnamefont {K.~K.}\ \bibnamefont {Berggren}}, \bibinfo
  {author} {\bibfnamefont {W.~D.}\ \bibnamefont {Oliver}}, \bibinfo {author}
  {\bibfnamefont {L.~S.}\ \bibnamefont {Levitov}},\ and\ \bibinfo {author}
  {\bibfnamefont {T.~P.}\ \bibnamefont {Orlando}},\ }\bibfield  {title}
  {\bibinfo {title} {Amplitude spectroscopy of a solid-state artificial atom},\
  }\href@noop {} {\bibfield  {journal} {\bibinfo  {journal} {Nature (London)}\
  }\textbf {\bibinfo {volume} {455}},\ \bibinfo {pages} {51} (\bibinfo {year}
  {2008})}\BibitemShut {NoStop}%
\bibitem [{\citenamefont {Stehlik}\ \emph {et~al.}(2012)\citenamefont
  {Stehlik}, \citenamefont {Dovzhenko}, \citenamefont {Petta}, \citenamefont
  {Johansson}, \citenamefont {Nori}, \citenamefont {Lu},\ and\ \citenamefont
  {Gossard}}]{StehlikPRB12}%
  \BibitemOpen
  \bibfield  {author} {\bibinfo {author} {\bibfnamefont {J.}~\bibnamefont
  {Stehlik}}, \bibinfo {author} {\bibfnamefont {Y.}~\bibnamefont {Dovzhenko}},
  \bibinfo {author} {\bibfnamefont {J.~R.}\ \bibnamefont {Petta}}, \bibinfo
  {author} {\bibfnamefont {J.~R.}\ \bibnamefont {Johansson}}, \bibinfo {author}
  {\bibfnamefont {F.}~\bibnamefont {Nori}}, \bibinfo {author} {\bibfnamefont
  {H.}~\bibnamefont {Lu}},\ and\ \bibinfo {author} {\bibfnamefont {A.~C.}\
  \bibnamefont {Gossard}},\ }\bibfield  {title} {\bibinfo {title}
  {{L}andau-{Z}ener-{S}t\"uckelberg interferometry of a single electron charge
  qubit},\ }\href {https://doi.org/10.1103/PhysRevB.86.121303} {\bibfield
  {journal} {\bibinfo  {journal} {Phys. Rev. B}\ }\textbf {\bibinfo {volume}
  {86}},\ \bibinfo {pages} {121303(R)} (\bibinfo {year} {2012})}\BibitemShut
  {NoStop}%
\bibitem [{\citenamefont {Forster}\ \emph {et~al.}(2014)\citenamefont
  {Forster}, \citenamefont {Petersen}, \citenamefont {Manus}, \citenamefont
  {H{\"a}nggi}, \citenamefont {Schuh}, \citenamefont {Wegscheider},
  \citenamefont {Kohler},\ and\ \citenamefont {Ludwig}}]{ForsterPRL14}%
  \BibitemOpen
  \bibfield  {author} {\bibinfo {author} {\bibfnamefont {F.}~\bibnamefont
  {Forster}}, \bibinfo {author} {\bibfnamefont {G.}~\bibnamefont {Petersen}},
  \bibinfo {author} {\bibfnamefont {S.}~\bibnamefont {Manus}}, \bibinfo
  {author} {\bibfnamefont {P.}~\bibnamefont {H{\"a}nggi}}, \bibinfo {author}
  {\bibfnamefont {D.}~\bibnamefont {Schuh}}, \bibinfo {author} {\bibfnamefont
  {W.}~\bibnamefont {Wegscheider}}, \bibinfo {author} {\bibfnamefont
  {S.}~\bibnamefont {Kohler}},\ and\ \bibinfo {author} {\bibfnamefont
  {S.}~\bibnamefont {Ludwig}},\ }\bibfield  {title} {\bibinfo {title}
  {{C}haracterization of qubit dephasing by
  {L}andau-{Z}ener-{S}t{\"u}ckelberg-{M}ajorana interferometry},\ }\href
  {https://doi.org/10.1103/PhysRevLett.112.116803} {\bibfield  {journal}
  {\bibinfo  {journal} {Phys. Rev. Lett.}\ }\textbf {\bibinfo {volume} {112}},\
  \bibinfo {pages} {116803} (\bibinfo {year} {2014})}\BibitemShut {NoStop}%
\bibitem [{\citenamefont {Forster}\ \emph {et~al.}(2015)\citenamefont
  {Forster}, \citenamefont {M\"uhlbacher}, \citenamefont {Blattmann},
  \citenamefont {Schuh}, \citenamefont {Wegscheider}, \citenamefont {Ludwig},\
  and\ \citenamefont {Kohler}}]{ForsterPRB15b}%
  \BibitemOpen
  \bibfield  {author} {\bibinfo {author} {\bibfnamefont {F.}~\bibnamefont
  {Forster}}, \bibinfo {author} {\bibfnamefont {M.}~\bibnamefont
  {M\"uhlbacher}}, \bibinfo {author} {\bibfnamefont {R.}~\bibnamefont
  {Blattmann}}, \bibinfo {author} {\bibfnamefont {D.}~\bibnamefont {Schuh}},
  \bibinfo {author} {\bibfnamefont {W.}~\bibnamefont {Wegscheider}}, \bibinfo
  {author} {\bibfnamefont {S.}~\bibnamefont {Ludwig}},\ and\ \bibinfo {author}
  {\bibfnamefont {S.}~\bibnamefont {Kohler}},\ }\bibfield  {title} {\bibinfo
  {title} {{L}andau-{Z}ener interference at bichromatic driving},\ }\href
  {https://doi.org/10.1103/PhysRevB.92.245422} {\bibfield  {journal} {\bibinfo
  {journal} {Phys. Rev. B}\ }\textbf {\bibinfo {volume} {92}},\ \bibinfo
  {pages} {245422} (\bibinfo {year} {2015})}\BibitemShut {NoStop}%
\bibitem [{\citenamefont {Heinrich}\ \emph {et~al.}(2021)\citenamefont
  {Heinrich}, \citenamefont {Oliver}, \citenamefont {Vandersypen},
  \citenamefont {Ardavan}, \citenamefont {Sessoli}, \citenamefont {Loss},
  \citenamefont {Bleszynski~Jayich}, \citenamefont {Fernandez-Rossier},
  \citenamefont {Laucht},\ and\ \citenamefont {Morello}}]{HeinrichNN21}%
  \BibitemOpen
  \bibfield  {author} {\bibinfo {author} {\bibfnamefont {A.~J.}\ \bibnamefont
  {Heinrich}}, \bibinfo {author} {\bibfnamefont {W.~D.}\ \bibnamefont
  {Oliver}}, \bibinfo {author} {\bibfnamefont {L.~M.~K.}\ \bibnamefont
  {Vandersypen}}, \bibinfo {author} {\bibfnamefont {A.}~\bibnamefont
  {Ardavan}}, \bibinfo {author} {\bibfnamefont {R.}~\bibnamefont {Sessoli}},
  \bibinfo {author} {\bibfnamefont {D.}~\bibnamefont {Loss}}, \bibinfo {author}
  {\bibfnamefont {A.}~\bibnamefont {Bleszynski~Jayich}}, \bibinfo {author}
  {\bibfnamefont {J.}~\bibnamefont {Fernandez-Rossier}}, \bibinfo {author}
  {\bibfnamefont {A.}~\bibnamefont {Laucht}},\ and\ \bibinfo {author}
  {\bibfnamefont {A.}~\bibnamefont {Morello}},\ }\bibfield  {title} {\bibinfo
  {title} {Quantum-coherent nanoscience},\ }\href
  {https://doi.org/10.1038/s41565-021-00994-1} {\bibfield  {journal} {\bibinfo
  {journal} {Nature Nanotech.}\ }\textbf {\bibinfo {volume} {16}},\ \bibinfo
  {pages} {1318} (\bibinfo {year} {2021})}\BibitemShut {NoStop}%
\bibitem [{Note1()}]{Note1}%
  \BibitemOpen
  \bibinfo {note} {In the interaction term, we have neglected the small
  difference of the moments of interia. Moreover, the sign of $\varepsilon (t)$
  is chosen such that it matches the usual definition in the quantum mechanical
  two-level problem. It is positive for attractive interaction.}\BibitemShut
  {Stop}%
\bibitem [{\citenamefont {Shevchenko}\ \emph {et~al.}(2010)\citenamefont
  {Shevchenko}, \citenamefont {Ashhab},\ and\ \citenamefont
  {Nori}}]{ShevchenkoPR10}%
  \BibitemOpen
  \bibfield  {author} {\bibinfo {author} {\bibfnamefont {S.~N.}\ \bibnamefont
  {Shevchenko}}, \bibinfo {author} {\bibfnamefont {S.}~\bibnamefont {Ashhab}},\
  and\ \bibinfo {author} {\bibfnamefont {F.}~\bibnamefont {Nori}},\ }\bibfield
  {title} {\bibinfo {title} {{L}andau-{Z}ener-{S}t\"uckelberg interferometry},\
  }\href {https://doi.org/10.1016/j.physrep.2010.03.002} {\bibfield  {journal}
  {\bibinfo  {journal} {Phys. Rep.}\ }\textbf {\bibinfo {volume} {492}},\
  \bibinfo {pages} {1} (\bibinfo {year} {2010})}\BibitemShut {NoStop}%
\bibitem [{\citenamefont {Ivakhnenko}\ \emph {et~al.}(2023)\citenamefont
  {Ivakhnenko}, \citenamefont {Shevchenko},\ and\ \citenamefont
  {Nori}}]{IvakhnenkoPR23}%
  \BibitemOpen
  \bibfield  {author} {\bibinfo {author} {\bibfnamefont {O.~V.}\ \bibnamefont
  {Ivakhnenko}}, \bibinfo {author} {\bibfnamefont {S.~N.}\ \bibnamefont
  {Shevchenko}},\ and\ \bibinfo {author} {\bibfnamefont {F.}~\bibnamefont
  {Nori}},\ }\bibfield  {title} {\bibinfo {title} {Nonadiabatic
  {L}andau-{Z}ener-{S}t\"uckelberg-{M}ajorana transitions, dynamics, and
  interference},\ }\href
  {https://doi.org/https://doi.org/10.1016/j.physrep.2022.10.002} {\bibfield
  {journal} {\bibinfo  {journal} {Phys. Rep.}\ }\textbf {\bibinfo {volume}
  {995}},\ \bibinfo {pages} {1} (\bibinfo {year} {2023})}\BibitemShut {NoStop}%
\bibitem [{\citenamefont {Mathieu}(1868)}]{MathieuJMPA68}%
  \BibitemOpen
  \bibfield  {author} {\bibinfo {author} {\bibfnamefont {E.}~\bibnamefont
  {Mathieu}},\ }\bibfield  {title} {\bibinfo {title} {Memoire sur le mouvement
  vibratoire d'une membrane de forme elliptique},\ }\href
  {http://eudml.org/doc/234720} {\bibfield  {journal} {\bibinfo  {journal} {J.
  Math. Pures Appl.}\ }\textbf {\bibinfo {volume} {13}},\ \bibinfo {pages}
  {137} (\bibinfo {year} {1868})}\BibitemShut {NoStop}%
\bibitem [{\citenamefont {Saito}\ \emph {et~al.}(2006)\citenamefont {Saito},
  \citenamefont {Wubs}, \citenamefont {Kohler}, \citenamefont {H\"anggi},\ and\
  \citenamefont {Kayanuma}}]{SaitoEL06}%
  \BibitemOpen
  \bibfield  {author} {\bibinfo {author} {\bibfnamefont {K.}~\bibnamefont
  {Saito}}, \bibinfo {author} {\bibfnamefont {M.}~\bibnamefont {Wubs}},
  \bibinfo {author} {\bibfnamefont {S.}~\bibnamefont {Kohler}}, \bibinfo
  {author} {\bibfnamefont {P.}~\bibnamefont {H\"anggi}},\ and\ \bibinfo
  {author} {\bibfnamefont {Y.}~\bibnamefont {Kayanuma}},\ }\bibfield  {title}
  {\bibinfo {title} {Quantum state preparation in circuit {QED} via
  {L}andau-{Z}ener tunneling},\ }\href@noop {} {\bibfield  {journal} {\bibinfo
  {journal} {Europhys. Lett.}\ }\textbf {\bibinfo {volume} {76}},\ \bibinfo
  {pages} {22} (\bibinfo {year} {2006})}\BibitemShut {NoStop}%
\bibitem [{\citenamefont {Ribeiro}\ and\ \citenamefont
  {Burkard}(2009)}]{RibeiroPRL09}%
  \BibitemOpen
  \bibfield  {author} {\bibinfo {author} {\bibfnamefont {H.}~\bibnamefont
  {Ribeiro}}\ and\ \bibinfo {author} {\bibfnamefont {G.}~\bibnamefont
  {Burkard}},\ }\bibfield  {title} {\bibinfo {title} {Nuclear state preparation
  via {L}andau-{Z}ener-{S}t\"uckelberg transitions in double quantum dots},\
  }\href {https://doi.org/10.1103/PhysRevLett.102.216802} {\bibfield  {journal}
  {\bibinfo  {journal} {Phys. Rev. Lett.}\ }\textbf {\bibinfo {volume} {102}},\
  \bibinfo {pages} {216802} (\bibinfo {year} {2009})}\BibitemShut {NoStop}%
\bibitem [{\citenamefont {Schr\"odinger}(1926)}]{SchrodingerPR26}%
  \BibitemOpen
  \bibfield  {author} {\bibinfo {author} {\bibfnamefont {E.}~\bibnamefont
  {Schr\"odinger}},\ }\bibfield  {title} {\bibinfo {title} {An undulatory
  theory of the mechanics of atoms and molecules},\ }\href
  {https://doi.org/10.1103/PhysRev.28.1049} {\bibfield  {journal} {\bibinfo
  {journal} {Phys. Rev.}\ }\textbf {\bibinfo {volume} {28}},\ \bibinfo {pages}
  {1049} (\bibinfo {year} {1926})}\BibitemShut {NoStop}%
\bibitem [{\citenamefont {Berry}(1984)}]{BerryPRSA84}%
  \BibitemOpen
  \bibfield  {author} {\bibinfo {author} {\bibfnamefont {M.~V.}\ \bibnamefont
  {Berry}},\ }\bibfield  {title} {\bibinfo {title} {Quantal phase factors
  accompanying adiabatic changes},\ }\href
  {https://doi.org/10.1098/rspa.1984.0023} {\bibfield  {journal} {\bibinfo
  {journal} {Proc. R. Soc. A}\ }\textbf {\bibinfo {volume} {392}},\ \bibinfo
  {pages} {45} (\bibinfo {year} {1984})}\BibitemShut {NoStop}%
\bibitem [{\citenamefont {Mi}\ \emph {et~al.}(2018)\citenamefont {Mi},
  \citenamefont {Kohler},\ and\ \citenamefont {Petta}}]{MiPRB18}%
  \BibitemOpen
  \bibfield  {author} {\bibinfo {author} {\bibfnamefont {X.}~\bibnamefont
  {Mi}}, \bibinfo {author} {\bibfnamefont {S.}~\bibnamefont {Kohler}},\ and\
  \bibinfo {author} {\bibfnamefont {J.~R.}\ \bibnamefont {Petta}},\ }\bibfield
  {title} {\bibinfo {title} {{L}andau-{Z}ener interferometry of valley-orbit
  states in {Si/SiGe} double quantum dots},\ }\href
  {https://doi.org/10.1103/PhysRevB.98.161404} {\bibfield  {journal} {\bibinfo
  {journal} {Phys. Rev. B}\ }\textbf {\bibinfo {volume} {98}},\ \bibinfo
  {pages} {161404(R)} (\bibinfo {year} {2018})}\BibitemShut {NoStop}%
\bibitem [{\citenamefont {Shevchenko}\ \emph {et~al.}(2018)\citenamefont
  {Shevchenko}, \citenamefont {Ryzhov},\ and\ \citenamefont
  {Nori}}]{ShevchenkoPRB18}%
  \BibitemOpen
  \bibfield  {author} {\bibinfo {author} {\bibfnamefont {S.~N.}\ \bibnamefont
  {Shevchenko}}, \bibinfo {author} {\bibfnamefont {A.~I.}\ \bibnamefont
  {Ryzhov}},\ and\ \bibinfo {author} {\bibfnamefont {F.}~\bibnamefont {Nori}},\
  }\bibfield  {title} {\bibinfo {title} {Low-frequency spectroscopy for quantum
  multilevel systems},\ }\href {https://doi.org/10.1103/PhysRevB.98.195434}
  {\bibfield  {journal} {\bibinfo  {journal} {Phys. Rev. B}\ }\textbf {\bibinfo
  {volume} {98}},\ \bibinfo {pages} {195434} (\bibinfo {year}
  {2018})}\BibitemShut {NoStop}%
\bibitem [{\citenamefont {Menchon-Enrich}\ \emph {et~al.}(2016)\citenamefont
  {Menchon-Enrich}, \citenamefont {Benseny}, \citenamefont {Ahufinger},
  \citenamefont {Greentree}, \citenamefont {Busch},\ and\ \citenamefont
  {Mompart}}]{MenchonEnrichRPP16}%
  \BibitemOpen
  \bibfield  {author} {\bibinfo {author} {\bibfnamefont {R.}~\bibnamefont
  {Menchon-Enrich}}, \bibinfo {author} {\bibfnamefont {A.}~\bibnamefont
  {Benseny}}, \bibinfo {author} {\bibfnamefont {V.}~\bibnamefont {Ahufinger}},
  \bibinfo {author} {\bibfnamefont {A.~D.}\ \bibnamefont {Greentree}}, \bibinfo
  {author} {\bibfnamefont {T.}~\bibnamefont {Busch}},\ and\ \bibinfo {author}
  {\bibfnamefont {J.}~\bibnamefont {Mompart}},\ }\bibfield  {title} {\bibinfo
  {title} {Spatial adiabatic passage: a review of recent progress},\ }\href
  {https://doi.org/10.1088/0034-4885/79/7/074401} {\bibfield  {journal}
  {\bibinfo  {journal} {Rep. Prog. Phys.}\ }\textbf {\bibinfo {volume} {79}},\
  \bibinfo {pages} {074401} (\bibinfo {year} {2016})}\BibitemShut {NoStop}%
\bibitem [{\citenamefont {Leggett}(2001)}]{LeggettRMP01}%
  \BibitemOpen
  \bibfield  {author} {\bibinfo {author} {\bibfnamefont {A.~J.}\ \bibnamefont
  {Leggett}},\ }\bibfield  {title} {\bibinfo {title} {Bose-{E}instein
  condensation in the alkali gases: Some fundamental concepts},\ }\href
  {https://doi.org/10.1103/RevModPhys.73.307} {\bibfield  {journal} {\bibinfo
  {journal} {Rev. Mod. Phys.}\ }\textbf {\bibinfo {volume} {73}},\ \bibinfo
  {pages} {307} (\bibinfo {year} {2001})}\BibitemShut {NoStop}%
\bibitem [{\citenamefont {Goldstein}\ \emph {et~al.}(2001)\citenamefont
  {Goldstein}, \citenamefont {Poole~Jr.},\ and\ \citenamefont
  {Safko}}]{Goldstein2001}%
  \BibitemOpen
  \bibfield  {author} {\bibinfo {author} {\bibfnamefont {H.}~\bibnamefont
  {Goldstein}}, \bibinfo {author} {\bibfnamefont {C.}~\bibnamefont
  {Poole~Jr.}},\ and\ \bibinfo {author} {\bibfnamefont {J.}~\bibnamefont
  {Safko}},\ }\href@noop {} {\emph {\bibinfo {title} {Classical Mechanics}}},\
  \bibinfo {edition} {3rd}\ ed.\ (\bibinfo  {publisher} {Pearson},\ \bibinfo
  {address} {San Francisco},\ \bibinfo {year} {2001})\BibitemShut {NoStop}%
\bibitem [{\citenamefont {Jackson}(1999)}]{Jackson1999}%
  \BibitemOpen
  \bibfield  {author} {\bibinfo {author} {\bibfnamefont {J.~D.}\ \bibnamefont
  {Jackson}},\ }\href@noop {} {\emph {\bibinfo {title} {Classical
  Electrodynamics}}},\ \bibinfo {edition} {3rd}\ ed.\ (\bibinfo  {publisher}
  {Wiley},\ \bibinfo {address} {New York},\ \bibinfo {year} {1999})\BibitemShut
  {NoStop}%
\bibitem [{\citenamefont {Cohen-Tannoudji}\ \emph {et~al.}(1992)\citenamefont
  {Cohen-Tannoudji}, \citenamefont {Dupont-Roc},\ and\ \citenamefont
  {Grynberg}}]{CohenTannoudji1992}%
  \BibitemOpen
  \bibfield  {author} {\bibinfo {author} {\bibfnamefont {C.}~\bibnamefont
  {Cohen-Tannoudji}}, \bibinfo {author} {\bibfnamefont {J.}~\bibnamefont
  {Dupont-Roc}},\ and\ \bibinfo {author} {\bibfnamefont {G.}~\bibnamefont
  {Grynberg}},\ }\href@noop {} {\emph {\bibinfo {title} {Atom photon
  interaction: Basic processes and applications}}}\ (\bibinfo  {publisher}
  {Wiley},\ \bibinfo {address} {New York},\ \bibinfo {year} {1992})\BibitemShut
  {NoStop}%
\bibitem [{\citenamefont {Oliver}\ \emph {et~al.}(2005)\citenamefont {Oliver},
  \citenamefont {Yu}, \citenamefont {Lee}, \citenamefont {Berggren},
  \citenamefont {Levitov},\ and\ \citenamefont {Orlando}}]{OliverS05}%
  \BibitemOpen
  \bibfield  {author} {\bibinfo {author} {\bibfnamefont {W.~D.}\ \bibnamefont
  {Oliver}}, \bibinfo {author} {\bibfnamefont {Y.}~\bibnamefont {Yu}}, \bibinfo
  {author} {\bibfnamefont {J.~C.}\ \bibnamefont {Lee}}, \bibinfo {author}
  {\bibfnamefont {K.~K.}\ \bibnamefont {Berggren}}, \bibinfo {author}
  {\bibfnamefont {L.~S.}\ \bibnamefont {Levitov}},\ and\ \bibinfo {author}
  {\bibfnamefont {T.~P.}\ \bibnamefont {Orlando}},\ }\bibfield  {title}
  {\bibinfo {title} {Mach-zehnder interferometry in a strongly driven
  superconducting qubit},\ }\href {https://doi.org/10.1126/science.1119678}
  {\bibfield  {journal} {\bibinfo  {journal} {Science}\ }\textbf {\bibinfo
  {volume} {310}},\ \bibinfo {pages} {1653} (\bibinfo {year}
  {2005})}\BibitemShut {NoStop}%
\bibitem [{\citenamefont {Sillanp\"a\"a}\ \emph {et~al.}(2006)\citenamefont
  {Sillanp\"a\"a}, \citenamefont {Lehtinen}, \citenamefont {Paila},
  \citenamefont {Makhlin},\ and\ \citenamefont {Hakonen}}]{SillanpaaPRL06}%
  \BibitemOpen
  \bibfield  {author} {\bibinfo {author} {\bibfnamefont {M.}~\bibnamefont
  {Sillanp\"a\"a}}, \bibinfo {author} {\bibfnamefont {T.}~\bibnamefont
  {Lehtinen}}, \bibinfo {author} {\bibfnamefont {A.}~\bibnamefont {Paila}},
  \bibinfo {author} {\bibfnamefont {Y.}~\bibnamefont {Makhlin}},\ and\ \bibinfo
  {author} {\bibfnamefont {P.}~\bibnamefont {Hakonen}},\ }\bibfield  {title}
  {\bibinfo {title} {Continuous-time monitoring of {L}andau-{Z}ener
  interference in a {C}ooper-pair box},\ }\href
  {https://doi.org/10.1103/PhysRevLett.96.187002} {\bibfield  {journal}
  {\bibinfo  {journal} {Phys. Rev. Lett.}\ }\textbf {\bibinfo {volume} {96}},\
  \bibinfo {pages} {187002} (\bibinfo {year} {2006})}\BibitemShut {NoStop}%
\bibitem [{\citenamefont {Wilson}\ \emph {et~al.}(2007)\citenamefont {Wilson},
  \citenamefont {Duty}, \citenamefont {Persson}, \citenamefont {Sandberg},
  \citenamefont {Johansson},\ and\ \citenamefont {Delsing}}]{WilsonPRL07}%
  \BibitemOpen
  \bibfield  {author} {\bibinfo {author} {\bibfnamefont {C.~M.}\ \bibnamefont
  {Wilson}}, \bibinfo {author} {\bibfnamefont {T.}~\bibnamefont {Duty}},
  \bibinfo {author} {\bibfnamefont {F.}~\bibnamefont {Persson}}, \bibinfo
  {author} {\bibfnamefont {M.}~\bibnamefont {Sandberg}}, \bibinfo {author}
  {\bibfnamefont {G.}~\bibnamefont {Johansson}},\ and\ \bibinfo {author}
  {\bibfnamefont {P.}~\bibnamefont {Delsing}},\ }\bibfield  {title} {\bibinfo
  {title} {Coherence times of dressed states of a superconducting qubit under
  extreme driving},\ }\href {https://doi.org/10.1103/PhysRevLett.98.257003}
  {\bibfield  {journal} {\bibinfo  {journal} {Phys. Rev. Lett.}\ }\textbf
  {\bibinfo {volume} {98}},\ \bibinfo {pages} {257003} (\bibinfo {year}
  {2007})}\BibitemShut {NoStop}%
\bibitem [{\citenamefont {Izmalkov}\ \emph {et~al.}(2008)\citenamefont
  {Izmalkov}, \citenamefont {van~der Ploeg}, \citenamefont {Shevchenko},
  \citenamefont {Grajcar}, \citenamefont {Il'ichev}, \citenamefont {H\"ubner},
  \citenamefont {Omelyanchouk},\ and\ \citenamefont {Meyer}}]{IzmalkovPRL08}%
  \BibitemOpen
  \bibfield  {author} {\bibinfo {author} {\bibfnamefont {A.}~\bibnamefont
  {Izmalkov}}, \bibinfo {author} {\bibfnamefont {S.~H.~W.}\ \bibnamefont
  {van~der Ploeg}}, \bibinfo {author} {\bibfnamefont {S.~N.}\ \bibnamefont
  {Shevchenko}}, \bibinfo {author} {\bibfnamefont {M.}~\bibnamefont {Grajcar}},
  \bibinfo {author} {\bibfnamefont {E.}~\bibnamefont {Il'ichev}}, \bibinfo
  {author} {\bibfnamefont {U.}~\bibnamefont {H\"ubner}}, \bibinfo {author}
  {\bibfnamefont {A.~N.}\ \bibnamefont {Omelyanchouk}},\ and\ \bibinfo {author}
  {\bibfnamefont {H.-G.}\ \bibnamefont {Meyer}},\ }\bibfield  {title} {\bibinfo
  {title} {Consistency of ground state and spectroscopic measurements on flux
  qubits},\ }\href {https://doi.org/10.1103/PhysRevLett.101.017003} {\bibfield
  {journal} {\bibinfo  {journal} {Phys. Rev. Lett.}\ }\textbf {\bibinfo
  {volume} {101}},\ \bibinfo {pages} {017003} (\bibinfo {year}
  {2008})}\BibitemShut {NoStop}%
\bibitem [{\citenamefont {Li}\ \emph {et~al.}(2013)\citenamefont {Li},
  \citenamefont {Silveri}, \citenamefont {Kumar}, \citenamefont {Pirkkalainen},
  \citenamefont {Veps\"al\"ainen}, \citenamefont {Chien}, \citenamefont
  {Tuorila}, \citenamefont {Sillanp\"a\"a}, \citenamefont {Hakonen},
  \citenamefont {Thuneberg},\ and\ \citenamefont {Paraoanu}}]{LiNC13}%
  \BibitemOpen
  \bibfield  {author} {\bibinfo {author} {\bibfnamefont {J.}~\bibnamefont
  {Li}}, \bibinfo {author} {\bibfnamefont {M.~P.}\ \bibnamefont {Silveri}},
  \bibinfo {author} {\bibfnamefont {K.~S.}\ \bibnamefont {Kumar}}, \bibinfo
  {author} {\bibfnamefont {J.-M.}\ \bibnamefont {Pirkkalainen}}, \bibinfo
  {author} {\bibfnamefont {A.}~\bibnamefont {Veps\"al\"ainen}}, \bibinfo
  {author} {\bibfnamefont {W.~C.}\ \bibnamefont {Chien}}, \bibinfo {author}
  {\bibfnamefont {J.}~\bibnamefont {Tuorila}}, \bibinfo {author} {\bibfnamefont
  {M.~A.}\ \bibnamefont {Sillanp\"a\"a}}, \bibinfo {author} {\bibfnamefont
  {P.~J.}\ \bibnamefont {Hakonen}}, \bibinfo {author} {\bibfnamefont {E.~V.}\
  \bibnamefont {Thuneberg}},\ and\ \bibinfo {author} {\bibfnamefont {G.~S.}\
  \bibnamefont {Paraoanu}},\ }\bibfield  {title} {\bibinfo {title} {Motional
  averaging in a superconducting qubit},\ }\href@noop {} {\bibfield  {journal}
  {\bibinfo  {journal} {Nature Comm.}\ }\textbf {\bibinfo {volume} {4}},\
  \bibinfo {pages} {1420} (\bibinfo {year} {2013})}\BibitemShut {NoStop}%
\bibitem [{\citenamefont {Dupont-Ferrier}\ \emph {et~al.}(2013)\citenamefont
  {Dupont-Ferrier}, \citenamefont {Roche}, \citenamefont {Voisin},
  \citenamefont {Jehl}, \citenamefont {Wacquez}, \citenamefont {Vinet},
  \citenamefont {Sanquer},\ and\ \citenamefont
  {De~Franceschi}}]{DupontFerrierPRL13}%
  \BibitemOpen
  \bibfield  {author} {\bibinfo {author} {\bibfnamefont {E.}~\bibnamefont
  {Dupont-Ferrier}}, \bibinfo {author} {\bibfnamefont {B.}~\bibnamefont
  {Roche}}, \bibinfo {author} {\bibfnamefont {B.}~\bibnamefont {Voisin}},
  \bibinfo {author} {\bibfnamefont {X.}~\bibnamefont {Jehl}}, \bibinfo {author}
  {\bibfnamefont {R.}~\bibnamefont {Wacquez}}, \bibinfo {author} {\bibfnamefont
  {M.}~\bibnamefont {Vinet}}, \bibinfo {author} {\bibfnamefont
  {M.}~\bibnamefont {Sanquer}},\ and\ \bibinfo {author} {\bibfnamefont
  {S.}~\bibnamefont {De~Franceschi}},\ }\bibfield  {title} {\bibinfo {title}
  {Coherent coupling of two dopants in a silicon nanowire probed by
  {L}andau-{Z}ener-{S}t\"uckelberg interferometry},\ }\href
  {https://doi.org/10.1103/PhysRevLett.110.136802} {\bibfield  {journal}
  {\bibinfo  {journal} {Phys. Rev. Lett.}\ }\textbf {\bibinfo {volume} {110}},\
  \bibinfo {pages} {136802} (\bibinfo {year} {2013})}\BibitemShut {NoStop}%
\bibitem [{\citenamefont {Koski}\ \emph {et~al.}(2018)\citenamefont {Koski},
  \citenamefont {Landig}, \citenamefont {P\'alyi}, \citenamefont {Scarlino},
  \citenamefont {Reichl}, \citenamefont {Wegscheider}, \citenamefont {Burkard},
  \citenamefont {Wallraff}, \citenamefont {Ensslin},\ and\ \citenamefont
  {Ihn}}]{KoskiPRL18}%
  \BibitemOpen
  \bibfield  {author} {\bibinfo {author} {\bibfnamefont {J.~V.}\ \bibnamefont
  {Koski}}, \bibinfo {author} {\bibfnamefont {A.~J.}\ \bibnamefont {Landig}},
  \bibinfo {author} {\bibfnamefont {A.}~\bibnamefont {P\'alyi}}, \bibinfo
  {author} {\bibfnamefont {P.}~\bibnamefont {Scarlino}}, \bibinfo {author}
  {\bibfnamefont {C.}~\bibnamefont {Reichl}}, \bibinfo {author} {\bibfnamefont
  {W.}~\bibnamefont {Wegscheider}}, \bibinfo {author} {\bibfnamefont
  {G.}~\bibnamefont {Burkard}}, \bibinfo {author} {\bibfnamefont
  {A.}~\bibnamefont {Wallraff}}, \bibinfo {author} {\bibfnamefont
  {K.}~\bibnamefont {Ensslin}},\ and\ \bibinfo {author} {\bibfnamefont
  {T.}~\bibnamefont {Ihn}},\ }\bibfield  {title} {\bibinfo {title} {Floquet
  spectroscopy of a strongly driven quantum dot charge qubit with a microwave
  resonator},\ }\href {https://doi.org/10.1103/PhysRevLett.121.043603}
  {\bibfield  {journal} {\bibinfo  {journal} {Phys. Rev. Lett.}\ }\textbf
  {\bibinfo {volume} {121}},\ \bibinfo {pages} {043603} (\bibinfo {year}
  {2018})}\BibitemShut {NoStop}%
\bibitem [{\citenamefont {Chen}\ \emph {et~al.}(2021)\citenamefont {Chen},
  \citenamefont {Wang}, \citenamefont {Kohler}, \citenamefont {Kang},
  \citenamefont {Lin}, \citenamefont {Gu}, \citenamefont {Li}, \citenamefont
  {Guo}, \citenamefont {Hu}, \citenamefont {Jiang}, \citenamefont {Cao},\ and\
  \citenamefont {Guo}}]{ChenPRB21}%
  \BibitemOpen
  \bibfield  {author} {\bibinfo {author} {\bibfnamefont {M.-B.}\ \bibnamefont
  {Chen}}, \bibinfo {author} {\bibfnamefont {B.-C.}\ \bibnamefont {Wang}},
  \bibinfo {author} {\bibfnamefont {S.}~\bibnamefont {Kohler}}, \bibinfo
  {author} {\bibfnamefont {Y.}~\bibnamefont {Kang}}, \bibinfo {author}
  {\bibfnamefont {T.}~\bibnamefont {Lin}}, \bibinfo {author} {\bibfnamefont
  {S.-S.}\ \bibnamefont {Gu}}, \bibinfo {author} {\bibfnamefont {H.-O.}\
  \bibnamefont {Li}}, \bibinfo {author} {\bibfnamefont {G.-C.}\ \bibnamefont
  {Guo}}, \bibinfo {author} {\bibfnamefont {X.}~\bibnamefont {Hu}}, \bibinfo
  {author} {\bibfnamefont {H.-W.}\ \bibnamefont {Jiang}}, \bibinfo {author}
  {\bibfnamefont {G.}~\bibnamefont {Cao}},\ and\ \bibinfo {author}
  {\bibfnamefont {G.-P.}\ \bibnamefont {Guo}},\ }\bibfield  {title} {\bibinfo
  {title} {Floquet state depletion in ac-driven circuit {QED}},\ }\href
  {https://doi.org/10.1103/PhysRevB.103.205428} {\bibfield  {journal} {\bibinfo
   {journal} {Phys. Rev. B}\ }\textbf {\bibinfo {volume} {103}},\ \bibinfo
  {pages} {205428} (\bibinfo {year} {2021})}\BibitemShut {NoStop}%
\end{thebibliography}

%

\end{document}